\documentclass[useAMS,usenatbib]{mn2e} 
\usepackage[switch]{lineno}
\usepackage{graphicx}  
\usepackage{amssymb}
\usepackage[usenames]{color}
\usepackage[T1]{fontenc}
\usepackage{ae,aecompl}

\usepackage{amsmath}	
\usepackage{subfigure}
\usepackage{listings}

\usepackage{xspace}
\usepackage{url}

\usepackage{times}
\usepackage[english]{babel}
\usepackage{mathrsfs}

\usepackage{threeparttable}
\usepackage{booktabs}

\newcommand{\ie}{i.\,e.} 
\newcommand{\eg}{e.\,g.} 

\usepackage{xcolor}
\definecolor{green}{HTML}{33CC33}
\definecolor{red}{HTML}{FF3300}
\definecolor{blue}{HTML}{3333FF}

\usepackage[format=plain,labelformat=default, labelsep=endash, labelfont=bf, margin=0pt, tableposition=top, font={footnotesize}]{caption}
\usepackage[colorlinks=true,citecolor=blue,urlcolor=green]{hyperref}
\usepackage{twoopt}
\usepackage[varg]{txfonts}
\usepackage[normalem]{ulem}

\renewcommand{\eqref}[1]{Equation~\ref{#1}}
\newcommand{\fref}[1]{Figure~\ref{#1}}
\newcommand{\tref}[1]{Table~\ref{#1}}
\newcommand{\sref}[1]{Section~\ref{#1}}

\newcommand{\pipe}{\ensuremath{\rm K2P^2}\xspace}
\newcommand{\targ}{\ensuremath{\rm E167}\xspace}
\newcommand{\targt}{\ensuremath{\rm E243}\xspace}
\newcommand{\numax}{\ensuremath{\nu_{\rm max}}\xspace}
\newcommand{\dnu}{\ensuremath{\Delta\nu}\xspace}
\newcommand{\kp}{\emph{Kepler}\xspace}

\newcommand{\teff}{\ensuremath{T_{\rm eff}}\xspace}
\newcommand{\feh}{\ensuremath{\rm [Fe/H]}\xspace}
\newcommand{\vsini}{\ensuremath{v\sin\, i_{\star}}\xspace}
\newcommand{\newarcsec}{\(\stackrel{\:''}{\textstyle.}\)}

\numberwithin{equation}{section}

\makeatletter
\def\maketag@@@#1{\hbox{\m@th\normalfont\normalsize#1}}
\makeatother

\bibpunct{(}{)}{;}{a}{}{,} 

\makeatletter
\newcommand*\mysize{%
  \@setfontsize\mysize{5.7}{8.0}%
}
\makeatother

\makeatletter
\newcommand*\tabsize{%
  \@setfontsize\tabsize{7.}{8.0}%
}
\makeatother

\title[Asteroseismology of Hyades stars]{Asteroseismology of the Hyades with K2: first detection of main-sequence solar-like oscillations in an open cluster}

\author[M. N. Lund et al.]{Mikkel~N.~Lund$^{1,2}$\thanks{E-mail: \href{mailto:lundm@bison.ph.bham.ac.uk}{lundm@bison.ph.bham.ac.uk}},
Sarbani~Basu$^{3}$,
V{\'{\i}}ctor~Silva~Aguirre$^{2}$,
William~J.~Chaplin$^{1,2}$,\newauthor 
Aldo~M.~Serenelli$^{4}$,
Rafael~A.~Garc{\'{\i}}a$^{5}$,
David~W.~Latham$^{6}$,
Luca~Casagrande$^{7}$,
\newauthor Allyson Bieryla$^{6}$,
Guy~R.~Davies$^{1,2}$,
Lucas~S.~Viani$^{3}$,
Lars~A.~Buchhave$^{8}$,
Andrea~Miglio$^{1,2}$,
\newauthor 
David~R.~Soderblom$^{9}$,
Jeff~A.~Valenti$^{9}$,
Robert~P.~Stefanik$^{6}$,
and Rasmus~Handberg$^{2}$\vspace*{0.5em}
\\ 
$^1$School of Physics and Astronomy, University of Birmingham, Edgbaston, Birmingham, B15 2TT, UK\\
$^2$Stellar Astrophysics Centre, Department of Physics and Astronomy, Aarhus University, Ny Munkegade 120, DK-8000 Aarhus C, Denmark\\
$^3$Department of Astronomy, Yale University, PO Box 208101, New Haven, CT 06520-8101, USA\\
$^4$Institute of Space Sciences (CSIC-IEEC), Campus UAB, Carrer de Can Magrans, s/n E-08193 Cerdanyola del Vall\`{e}s (Barcelona), Spain\\
$^5$Laboratoire AIM, CEA/DRF - CNRS - Univ. Paris Diderot - IRFU/SAp, Centre de Saclay, 91191 Gifsur-Yvette Cedex, France\\
$^6$Harvard-Smithsonian Center for Astrophysics, 60 Garden Street Cambridge, MA 02138 USA\\
$^7$Research School of Astronomy and Astrophysics, Mount Stromlo Observatory, The Australian National University, ACT 2611, Australia\\
$^8$Centre for Star and Planet Formation, Natural History Museum of Denmark \& Niels Bohr Institute, University of Copenhagen,\\ \O ster Voldgade 5-7, DK-1350 Copenhagen K, Denmark\\
$^9$Space Telescope Science Institute, 3700 San Martin Drive, Baltimore, MD 21218, USA\\
}
   
\begin{document}

\vspace{-10cm}
\date{Accepted 2016 August 24. Received on 2016 August 23; in original form 2016 April 7}
\pagerange{\pageref{firstpage}--\pageref{lastpage}} \pubyear{2016}

\maketitle

\label{firstpage}
\vspace{-3cm}
\begin{abstract}

The Hyades open cluster was targeted during Campaign~4 (C4) of the
NASA K2 mission, and short-cadence data were collected on a number of
cool main-sequence stars. Here, we report results on two F-type stars
that show detectable oscillations of a quality that allows
asteroseismic analyses to be performed. These are the first ever
detections of solar-like oscillations in main-sequence stars in an
open cluster.

\end{abstract}


\section{Introduction}

The Hyades is one of the youngest and closest open clusters to our solar system; its close proximity of only ${\sim}47$ pc means that it has been extensively studied, and serves as an important benchmark for distances in our Galaxy \citep[see][for a review]{1998A&A...331...81P}. Because of its youth (with an isochrone-based age estimated to be around $550-625$ Myr) it contains many rapidly rotating stars whose rotation rates can be readily determined, hence it is commonly used as an anchor in calibrating gyrochronology relations which link rotation rates to stellar ages.

Asteroseismology --- the study of stellar oscillations --- offers independent measures of stellar properties. Results from the \kp mission have shown the power of asteroseismology in relation to characterisation and age dating of both field and cluster stars \citep[][]{2010PASP..122..131G,2011ApJ...729L..10B,2011ApJ...739...13S,2012MNRAS.419.2077M,2014ApJS..210....1C,2015MNRAS.452.2127S}. 
Regrettably, the nominal \kp mission did not observe nearby clusters, but K2, the repurposed \kp mission \citep[][]{2014PASP..126..398H}, will allow us to study many interesting clusters.
In this Letter we present the first asteroseismic analysis of main-sequence (MS) stars in the Hyades, specifically two MS solar-like oscillators. 
We note in passing that White et al. (in prep.) from K2 observations, and \citet[][]{2015A&A...573A.138B} from a ground-based campaign, have detected oscillations in three Hyades red-giants.  

The development of the paper proceeds as follows: In \sref{sec:data} we discuss the reduction of K2 data. We also describe the other known properties of the targets and present a set of new radial velocity data designed to confirm cluster membership and identify short-period binaries, and introduce in \sref{sec:atmos} the spectroscopic analysis of the stars. In \sref{sec:pb} we discuss how the asteroseimic data were used to determine stellar parameters. In \sref{sec:rot} we present our analysis of the signatures of rotation; the seismic modelling is presented in \sref{sec:model}, with distance estimates using the asteroseismic properties being compared to other distance indicators in \sref{sec:dist}. We end with a discussion of our findings in \sref{sec:dis}.


\section{Data}
\label{sec:data}

The Hyades open cluster, seen in the constellation of Taurus, was observed in short-cadence (SC; $\Delta {t\approx 1}\rm \ min$) during Campaign 4 (C4) of the K2 mission. SC data were collected for a total of 14 targets in the Hyades region.

Light curves were extracted from background-corrected pixel data\footnote{downloaded from the KASOC database; \url{www.kasoc.phys.au.dk}} using the \pipe pipeline \citep[][]{2015ApJ...806...30L}. Briefly, \pipe defines pixel masks for targets in a given frame by using an unsupervised clustering algorithm on pixels above a given flux threshold. Subsequently, an image segmentation algorithm is run on each pixel-cluster to adjust the pixel mask should two or more targets happen to fall within it. The light curves were rectified using a modified version of the KASOC filter \citep[][]{2014MNRAS.445.2698H} to remove trends from the apparent motion of the targets on the CCD and other instrumental signatures. Power density spectra were created using a least-squares sine-wave fitting method, normalised by the \textsc{rms}-scaled version of Parseval's theorem \citep[see][]{1992PhDT.......208K,1995A&A...301..123F}.

\begin{figure*}
    \centering
    \begin{subfigure}
        \centering
        \includegraphics[width=\columnwidth]{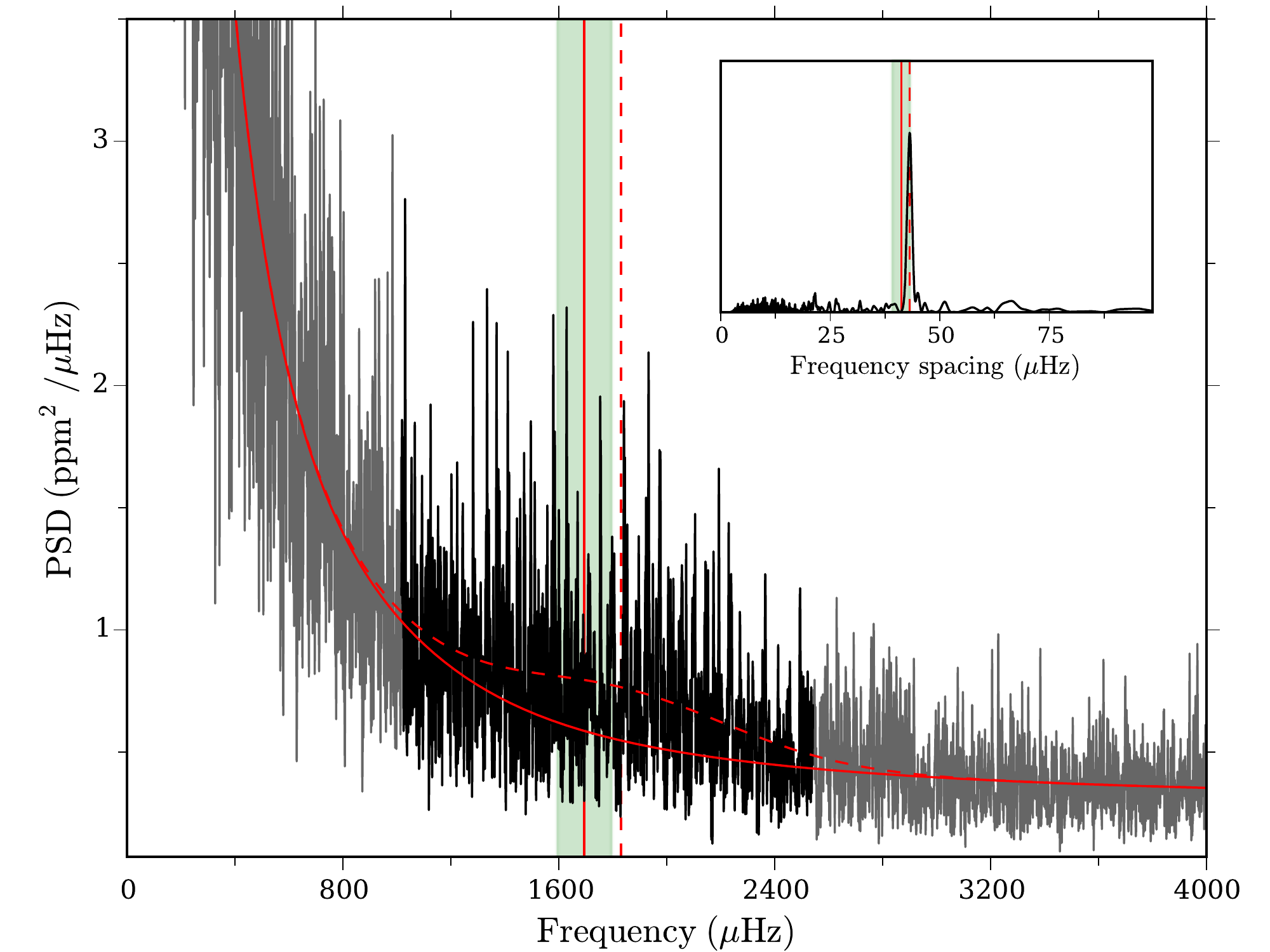}
    \end{subfigure}%
    \hfill 
    \begin{subfigure}
        \centering
        \includegraphics[width=\columnwidth]{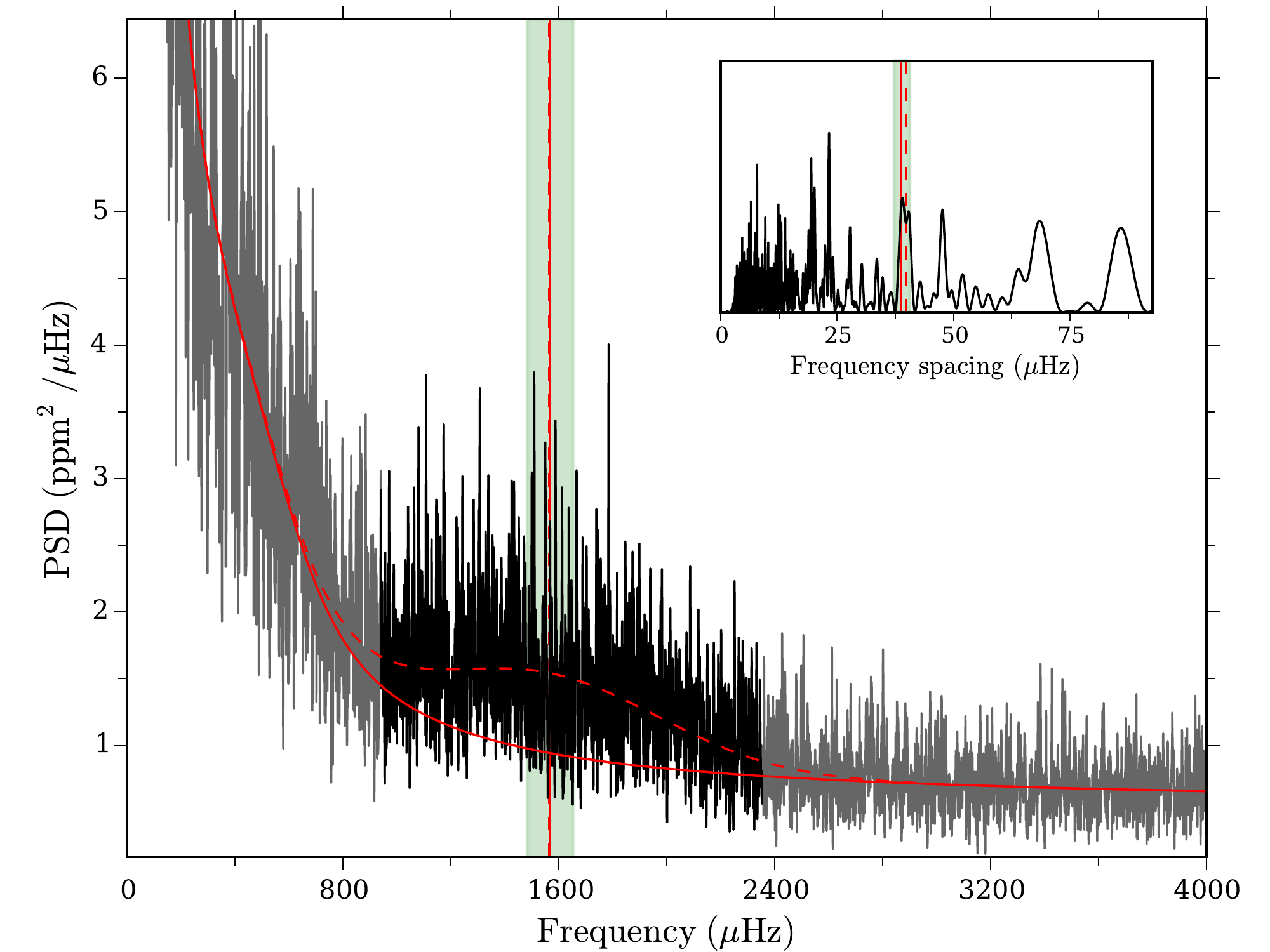}
    \end{subfigure}
    \caption{Power density spectra of \targ (left) and \targt (right), smoothed by a $3\,\rm \mu Hz$ Epanechnikov filter. The inserts show power-of-power spectra ($\rm PS\otimes PS$) from the region of the power spectra in black. The (full) vertical red lines indicate the spectroscopically estimated values for \numax of $1695\pm 100\, \rm \mu Hz$ (\targ) and $1568\pm86\, \rm \mu Hz$ (\targt), with uncertainties given by the shaded green region; the corresponding lines in the $\rm PS\otimes PS$ inserts give the expected values for $\dnu/2$ from these \numax estimates following \citet[][]{2011ApJ...743..143H}. Dashed vertical lines indicate the measured values for \numax and $\dnu$ (see \sref{sec:pb}). The full red lines show the obtained fits to the granulation backgrounds, with the red dashed line showing the fitted Gaussian envelope used to estimate \numax.}
\label{fig:ps}
\end{figure*}

We searched the power spectra of all observed stars for indications of seismic excess power --- two targets were identified, EPIC $210444167$ and $210499243$; from here on we will refer to these as \targ and \targt.
Based on proper motion and radial velocity studies by, \eg, \citet[][]{1991A&A...243..386S}, \citet[][]{1998A&A...331...81P}, and \citet[][]{2001A&A...367..111D} both targets are members of the Hyades. In \fref{fig:ps} we show the power spectra for the targets. The stars have spectral types F5 IV-V  \citep[\targ;][]{2001AJ....121.2148G} and F5 V \citep[\targt;][]{2010A&A...523A..71G}.

The star \targt has been studied before. \targt was specifically highlighted in \citet[][]{2001A&A...367..111D} for lying above the Hyades main-sequence $(\Delta V \sim 0.07\,\rm mag)$, and it was speculated if stellar variability or activity could be responsible for this, but at that time a good estimate of the rotational velocity was unavailable.
In the Catalog of Components of Double and Multiple stars \citep[CCDM;][]{2002yCat.1274....0D}\footnote{CCDM J04158+1525A/WDS J04158+1524A (WDS; The Washington Double Star Catalog, \citealt[][]{2001AJ....122.3466M})} the star (A) is listed with two secondary components (B and C), both with magnitudes in the range $V{\sim}11.5$ --- this multiplicity would give a change of $(\Delta V \sim 0.02\,\rm mag)$. We note, however, that the components listed in the CCDM are at very large separations of 137.5\arcsec (AB) and 151.4\arcsec (AC), corresponding to ${\sim}35$ and ${\sim}39$ pixels on the \kp CCD. The B component (Ba) has itself a faint companion (Bb), and the C component is a spectroscopic binary. From proper motions, radial velocity (RV) data from the Harvard-Smithsonian Center for Astrophysics (CfA), and colours neither Ba, Bb, nor C are associated with \targt or the Hyades cluster. Neither of these targets fall within the assigned pixel mask of \targt, and the photometry for \targt A is thus unaffected by B and C. 
\targt was furthermore found to be a single system from an analysis of speckle images by \citet[][]{1998AJ....115.1972P}; this does not necessarily, however, rule out a very close companion within the 0\newarcsec05 confusion limit of the speckle analysis.

Radial velocities can provide additional constraints on the possibility that close unresolved companions are contaminating the light of \targ and \targt. 
Both stars have been monitored for more than 35 years using RV instruments at CfA, and both appear to be single-lined, with no direct evidence for light from a companion. The velocities for \targ appear to be constant, but there is suggestive evidence for acceleration in a long-period spectroscopic
orbit for \targt (\fref{fig:vel}). There is insufficient information to put a strong constraint on the possible light contamination from a faint companion to \targt, but a contribution of several percent
cannot be ruled out. Four instruments have been used for the CfA velocities reported here for the first time (see Appendix~\ref{sec:rv}):
three almost identical versions of the CfA Digital Speedometers \citep[][]{1992ASPC...32..110L} on the 1.5-m Wyeth Reflector at the Oak Ridge Observatory in the Town of Harvard,
Massachusetts, and on the MMT and 1.5-m Tillinghast Reflector at the Fred Lawrence Whipple Observatory on Mount Hopkins, Arizona; and more recently the Tillinghast
Reflector Echelle Spectrograph \citep[TRES;][]{2007RMxAC..28..129S, furesz_phd}, a modern fiber-fed CCD echelle spectrograph on Mount Hopkins. 
Both stars show substantial line broadening due to rotation, ${\sim}22\, \rm km\, s^{-1}$ for \targ and ${\sim}67\, \rm km\, s^{-1}$ for \targt, so the line profiles are heavily
oversampled at the instrumental resolutions of about $6.7\, \rm km\, s^{-1}$. Velocities were derived using the most appropriate rotationally broadened templates from the CfA library of
synthetic spectra and are reported here on the native CfA system, which is about $0.14\, \rm km\, s^{-1}$ more negative than the IAU system. Thus $0.14\, \rm km\, s^{-1}$ should be added to the velocities
reported in \tref{tab:rv} and plotted in \fref{fig:vel} to put them onto the IAU system.
Mean radial velocities in the IAU system of $38.80\pm 0.63\, \rm km\, s^{-1}$ (\targ) and $39.87\pm 1.83\, \rm km\, s^{-1}$ (\targt) are obtained, where the uncertainties are given by the root-mean-square (\textsc{rms}) values of the individual velocities. These mean velocities agree well with the mean radial velocity of $39.42\pm 0.36\, \rm km\, s^{-1}$ derived by \citet[][]{2002A&A...381..446M} for the Hyades from a moving-cluster analysis, and thus stipulate to the Hyades membership of the stars.

There are a few additional historical velocities for \targ and \targt in the literature, extending the time coverage to 82 and 101 years, respectively. Unfortunately the
precision for the earliest velocities is poor and the systematic offset of the zero points is not well established. The historical velocities do strengthen the impression that \targ has
been constant, and the velocity of \targt was lower 100 years ago.

\section{Atmospheric and stellar parameters}
\label{sec:atmos}

We have obtained spectroscopic parameters for the targets from several sources:
(1) Values are available from the Geneva-Copenhagen Survey \citep[GCS;][]{2004A&A...418..989N} in their re-derived version by \citet[][]{2011A&A...530A.138C}; (2) Spectroscopic data were collected using the TRES spectrograph on the 1.5-m Tillinghast telescope at the F.~L.~Whipple Observatory; atmospheric parameters were derived using the Stellar Parameter Classification pipeline \citep[SPC;][]{2012Natur.486..375B}. Following \citet[][]{2012ApJ...757..161T} we added in quadrature uncertainties of $59$ K and $0.062$ dex to the \teff and $\rm [M/H]$ from SPC; (3) We also estimated \teff using the Infra-Red Flux Method \citep[IRFM;][]{2014ApJ...787..110C}. This method also gives a measure of the stellar angular diameter $\theta$ which combined with the parallax provides an independent estimate of the stellar radius \citep[][]{2012ApJ...757...99S}. A reddening of $E(B-V)=0.003\pm0.002$ \citep[][]{1980AJ.....85..242T} was adopted for the IRFM derivation. Reddening was neglected in the derivation of the GCS values, but the low value for $E(B-V)$ has virtually no impact on the derived stellar parameters. Final SPC parameters were obtained after iterating with a $\log g$ fixed at the asteroseismic value determined from \numax and \teff from $g\propto \numax \sqrt{\teff}$, and with a fixed metallicity of $\rm [M/H]=0.164\pm 0.08$. The metallicity is obtained from the average of the recent spectroscopic analysis results of Hyades members by \citet[][]{2016MNRAS.457.3934L}. We note that the adopted value from \citet[][]{2016MNRAS.457.3934L} agrees well, within the adopted uncertainty of $0.08\rm\, dex$, with previous average estimates from, for instance, \citet[][]{1985A&A...146..249C,1988ApJ...332..410B,1990ApJ...351..467B,1998A&A...331...81P,2003AJ....125.3185P,2013PASJ...65...53T}; and \citet{2016A&A...585A..75D}. A fixed metallicity was adopted because the SPC pipeline has difficulties with stars with a value of \vsini as high as that inferred for \targt (Table~\ref{tab:params}); this is because high rotation leads to rotational broadening that might cause blending of lines. The overall agreement between the different parameter sets does, however, lend credibility to the SPC values. Final parameters are given in Table~\ref{tab:params} --- \teff and $\rm [M/H]$ will serve as constraints in the asteroseismic modelling presented in \sref{sec:model}.

\begin{table*}
\tabsize
\centering
\caption{\scriptsize Spectroscopic parameters and common identifications for Hyades targets with detected oscillations. We give values obtained from the Stellar Parameter Classification pipeline \citep[SPC;][]{2012Natur.486..375B}, the Geneva-Copenhagen Survey \citep[GCS;][]{2004A&A...418..989N} in their re-derived version by \citet[][]{2011A&A...530A.138C}, and the InfraRed Flux Method \citep[IRFM; see][]{2014ApJ...787..110C}. Angular diameters ($\theta$) are from the IRFM. Systematic uncertainties of $59$ K and $0.062$ dex were added in quadrature to the SPC $T_{\rm eff}$ and [M/H] following \citet[][]{2012ApJ...757..161T}. We have highlighted in bold face the measured seismic values of \dnu and \numax. SPC values were iterated with a $\log g$ fixed to the seismic estimate and a fixed metallicity of $\rm [M/H]=0.164$ \citep[][]{2016MNRAS.457.3934L}.}
\label{tab:params}
\begin{tabular}{lllcccclcccc}
\toprule
EPIC & HIP & HD & Kp    & $\theta$ &  \numax          & \dnu  &  Source & \teff & [M/H]          & $\log g$       & $v\sin\, i_{\star}$  \\
     &     & & (mag)        & (mas)&  ($\rm \mu Hz$)  & ($\rm \mu Hz$) &  & (K)           &  (dex)         & ($\rm cgs$; dex)  & ($\rm km\ s^{-1}$)\\
\midrule
$210444167^{a}$  & 20357 & 27561 & 6.545  & $0.296\pm0.008$ &$\mathbf{1831\pm 47}$ & $\mathbf{86.2\pm 1.5}$ &  SPC  & $6761\pm 77$  & $0.164\pm 0.080$ & $4.24\pm 0.1$ & $22.0\pm 0.5$\\
              &       & &       &                &     && GCS & $6695\pm 102$ & $0.07$ & $4.15$ & \\
              &       & &       &                &     &&  IRFM & $6711\pm 81$ &  &  & \\              
$210499243^{b}$  & 19877 & 26911 & 6.264 & $0.331\pm0.010$ & $\mathbf{1564\pm 58}$ & $\mathbf{79.6\pm2.0}$& SPC & $6901\pm 77$ & $0.164\pm 0.080$ & $4.18\pm 0.1$ & $66.8\pm 0.5$ \\
              &       & &                    &   &   &&GCS  & $6765\pm 80$ & 0.17 &4.18 & \\
              &       & &                    &   &    && IRFM & $6771\pm 81$ &  &  & \\               
\bottomrule\\
\end{tabular}
\begin{tablenotes}
	\scriptsize\vspace{-1em}
    \item $^{a}$Also known as Cl Melotte 25 37, vB 37; $^{b}$also known as $48$ Tau, V$1099$ Tau, HR 1319, Cl Mellote 25 20, vB 20. 
  \end{tablenotes}
\end{table*}

From the spectroscopic parameters we can predict values for \numax using scaling relations \citep[][]{1995A&A...293...87K}. Masses were estimated from the IRFM \teff via the Hyades isochrone from \citet[][]{2004ApJ...600..946P}, radii from $L$ and \teff, and using ${\nu_{\rm max,\sun} = 3090\pm30\, \rm \mu Hz}$, and $T_{\rm eff, \sun} = 5777$ K \citep[][]{2011ApJ...743..143H,2014ApJS..210....1C}. In \fref{fig:ps} the estimates are seen to agree well with the seismic power excess and the measured values of \numax.
For the above prediction we estimated luminosities from kinematically improved parallaxes by \citet[][]{2002A&A...381..446M} and $V$-band magnitudes from \citet[][]{2006AJ....132..111J}. 
The relations of \citet[][]{1996ApJ...469..355F} as presented in \citet[][]{2010AJ....140.1158T} were used for the bolometric correction. 
Such a comparison is valuable, because it allows us a check of our predictions against the estimated seismic observables, and thus our ability to securely propose targets for future K2 campaigns.


\section{Analysis}



\subsection{Asteroseismic parameter estimation}
\label{sec:pb}
We first determined the global asteroseismic properties $\Delta\nu$ and \numax. Here $\Delta\nu$ is defined as the frequency spacing between consecutive radial orders ($n$) of modes with a given angular degree ($l$), and \numax as the frequency where the modes show their maximum amplitudes. 
To estimate \numax we fit the stellar noise background following the procedure described in \citet[][]{2014A&A...570A..54L}. For the background we adopt a model given by a sum of power laws with free exponents, one for each phenomenon contributing to the background \citep[see, \eg,][and references therein]{2014A&A...570A..41K}, and include a Gaussian envelope to account for the power excess from oscillations.
The obtained background fits are shown in \fref{fig:ps}.
We estimate \dnu from fit of a squared Gaussian function including a background to a narrow range of the power-of-power spectrum ($\rm PS\otimes PS$) centred on the $\dnu/2$ peak. We note that the small frequency separation $\delta\nu_{02}$ --- given by the frequency difference between adjacent $l=0$ and $l=2$ modes as $\delta\nu_{02}=\nu_{n,0} - \nu_{n-1,2}$ --- could not be estimated from the data. See Table~\ref{tab:params} for extracted parameters. The extracted values for \numax and \dnu agree within uncertainties with the $\dnu\propto \beta\numax^{\alpha}$ scaling by \citet[][]{2011ApJ...743..143H}.

\begin{figure}
\centering
\includegraphics[width=\columnwidth]{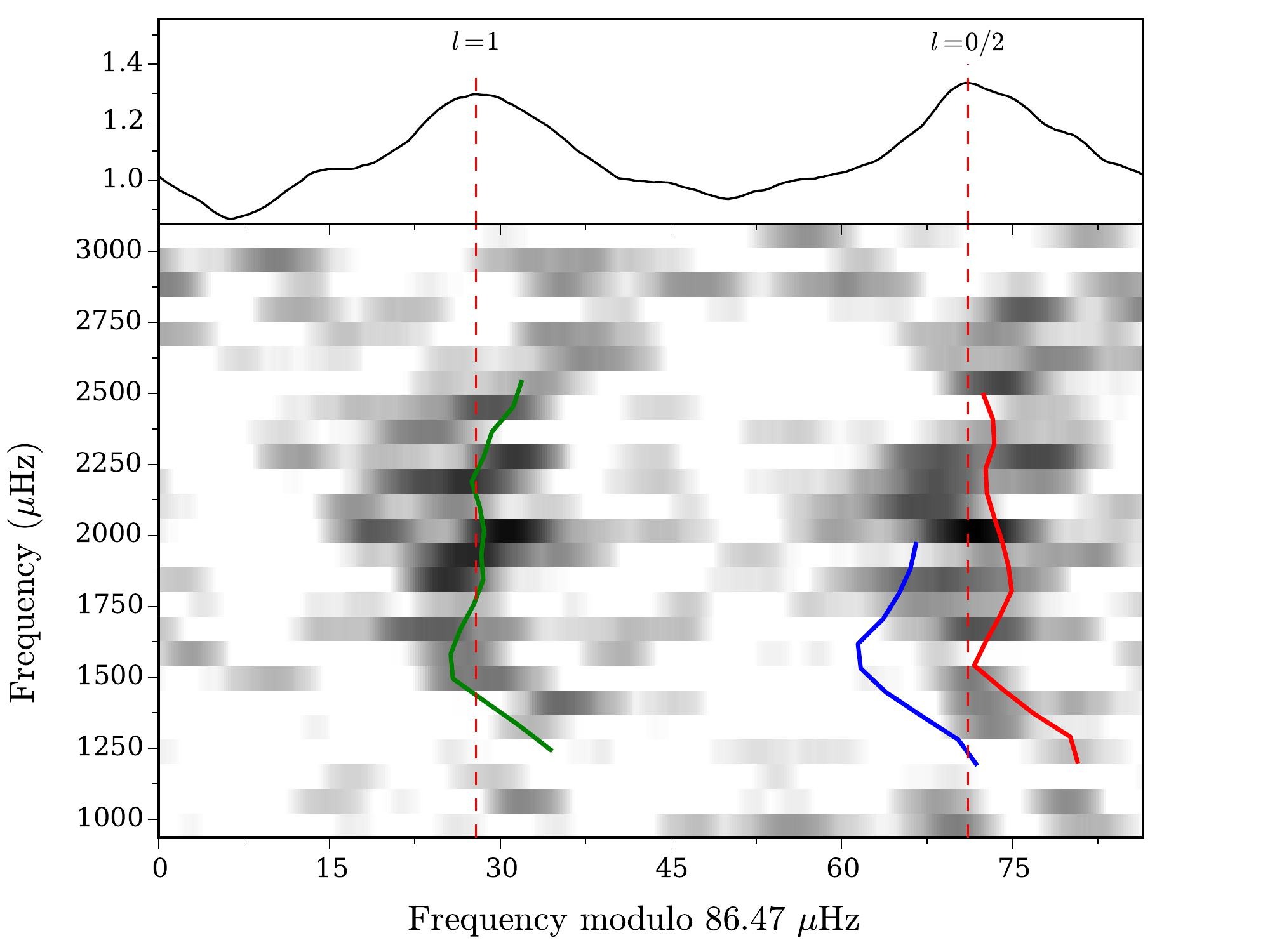}
\caption{Grey scale \'{e}chelle diagram for \targ, smoothed horizontally to a resolution of $\rm 10\, \mu Hz$. The lines show the frequency-ridges of KIC 3733735 (red $l=0$; green $l=1$; blue $l=2$) after multiplying by 0.9361 and forcing (by eye) the $\epsilon$ to match that of \targ by shifting the ridges by $\rm 6\, \mu Hz$. The top panel show the vertically collapsed \'{e}chelle giving the combined signal of the ridges. }
\label{fig:echelle}
\end{figure}

In \fref{fig:echelle} we show the background corrected \'{e}chelle diagram \citep[][]{1983SoPh...82...55G} for \targ, smoothed to a resolution of $\rm 10\, \mu Hz$. Over-plotted is the scaled \'{e}chelle diagram \citep[see][]{2010CoAst.161....3B} of frequencies for KIC 3733735\footnote{a.k.a. \textit{Shere-Khan} in the KASC working group 1 CATalogue.}, which in terms of fundamental parameters is similar to \targ, especially the similar age estimated at $800\pm400$ Myr \citep[][]{2014ApJS..210....1C}. It is noteworthy how well the structure in the ridges of KIC 3733735 match that of \targ, which indicates that the targets are indeed very similar. KIC 3733735, which is also a fast rotator, has been studied in relation to activity and rotation by \citet[][]{2014A&A...562A.124M} and Keifer et al. (2016; submitted). The ridge identification from this scaling matches that obtained using the $\epsilon$-method by \citet[][]{2011ApJ...742L...3W}.

The determination of $\Delta\nu$ for \targt is more uncertain than that for \targ, as also seen from the $\rm PS\otimes PS$ in \fref{fig:ps}.
We believe the reason for this can be found in the combination of the rotation rate, which for both stars is high, and stellar inclination --- as described in \sref{sec:rot} below, \targ is likely seen at a low inclination angle, while \targt seems to be observed edge-on. For \targ this would greatly decrease the visibility of rotationally split ($m\neq 0$) mode components, leaving with highest visibility the zonal ($m=0$) components \citep[][]{2003ApJ...589.1009G}. First of all, this would explain the distinguishable ridges in the \'{e}chelle diagram (\fref{fig:echelle}). Additionally, the value of $\rm\delta_{01} \approx 0\, \mu Hz$ obtained from the ridge averages shown in \fref{fig:echelle} explains the strong signal in the $\rm PS\otimes PS$ at $\Delta\nu/2$ --- with $\rm\delta_{01}$ given as the offset of $l = 1$ modes from the midpoint between the surrounding $l = 0$ modes, \ie, $\delta_{\rm 01}=  \tfrac{1}{2}(\nu_{n,0} + \nu_{n+1,0})-\nu_{n,1}$ \citep[see, \eg,][]{2011arXiv1107.1723B}. On the other hand, the $i_{\star}\approx 90^{\circ}$ configuration for \targt would maximise the rotational confusion of the power spectrum, and the difficulty in extracting $\Delta\nu$.
Concerning the estimation of \dnu we tested the effect of adding rotation on the $\rm PS\otimes PS$ and found that this had a negligible impact on the central position of the $\Delta\nu$ and $\Delta\nu/2$ peaks, and this both from an inclination of $i_{\star}=0^{\circ}$ and $90^{\circ}$. The main effect observed was a change in the relative heights between the peaks. This suggests that that $\rm PS\otimes PS$ provides a robust estimate of \dnu even in the case of high rotation, in agreement with \citet[][]{2009A&A...508..877M} who obtained good \dnu estimates for some fast rotators observed by CoRoT \citep[][]{2009A&A...506..245M}.
We checked for needed line-of-sight corrections to \numax following \citet[][]{2014MNRAS.445L..94D} --- For \targ this amounts to $\delta\nu_s\approx 0.23\,\rm \mu Hz$ at \numax; for \targt $\delta\nu_s\approx 0.19\,\rm \mu Hz$, both well within our adopted uncertainties on \numax.


\subsection{Rotation}
\label{sec:rot}
\begin{figure*}
    \centering
    \begin{subfigure}
        \centering
        \includegraphics[width=\columnwidth]{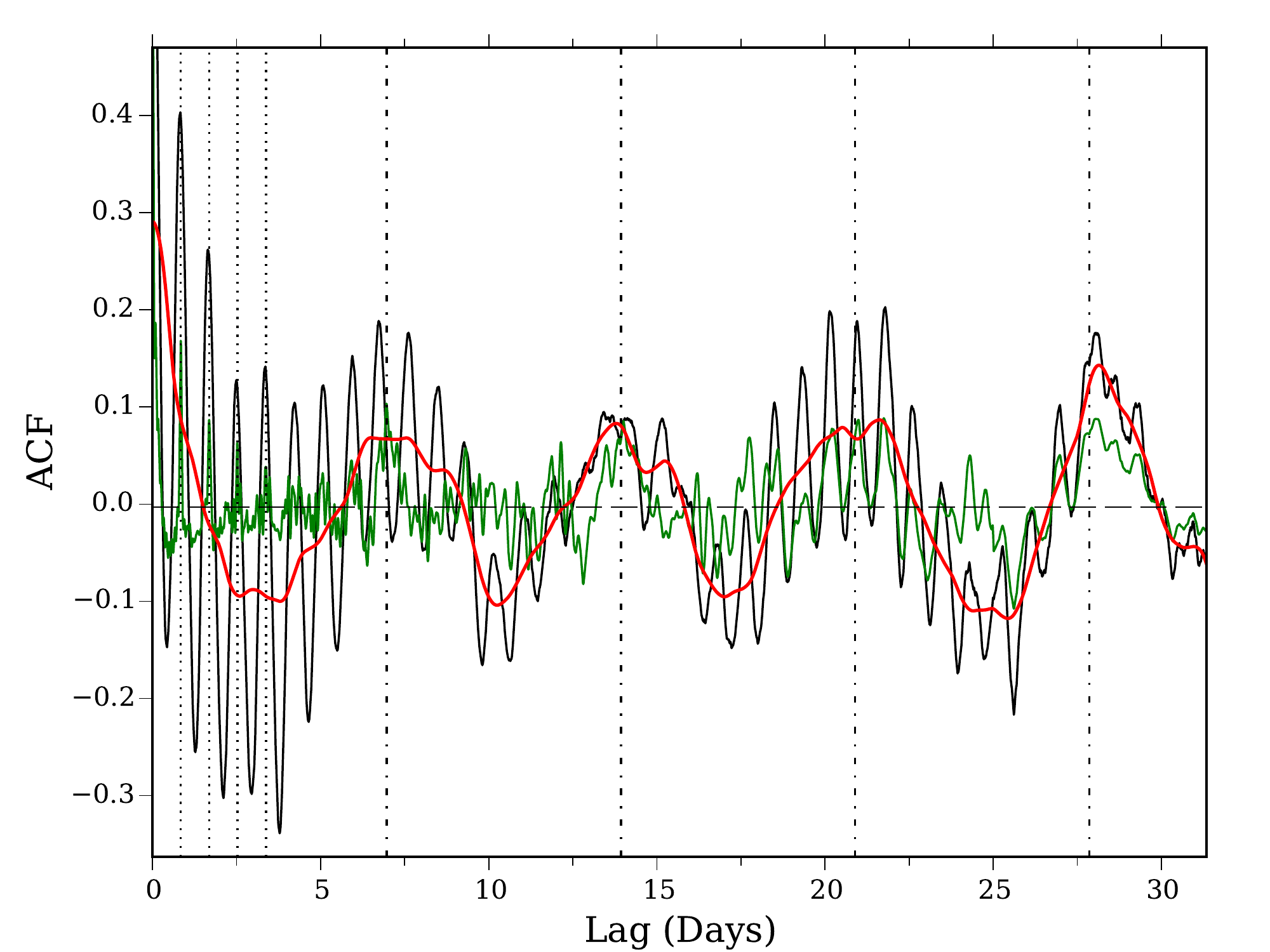}
    \end{subfigure}%
    \hfill
    \begin{subfigure}
        \centering
        \includegraphics[width=\columnwidth]{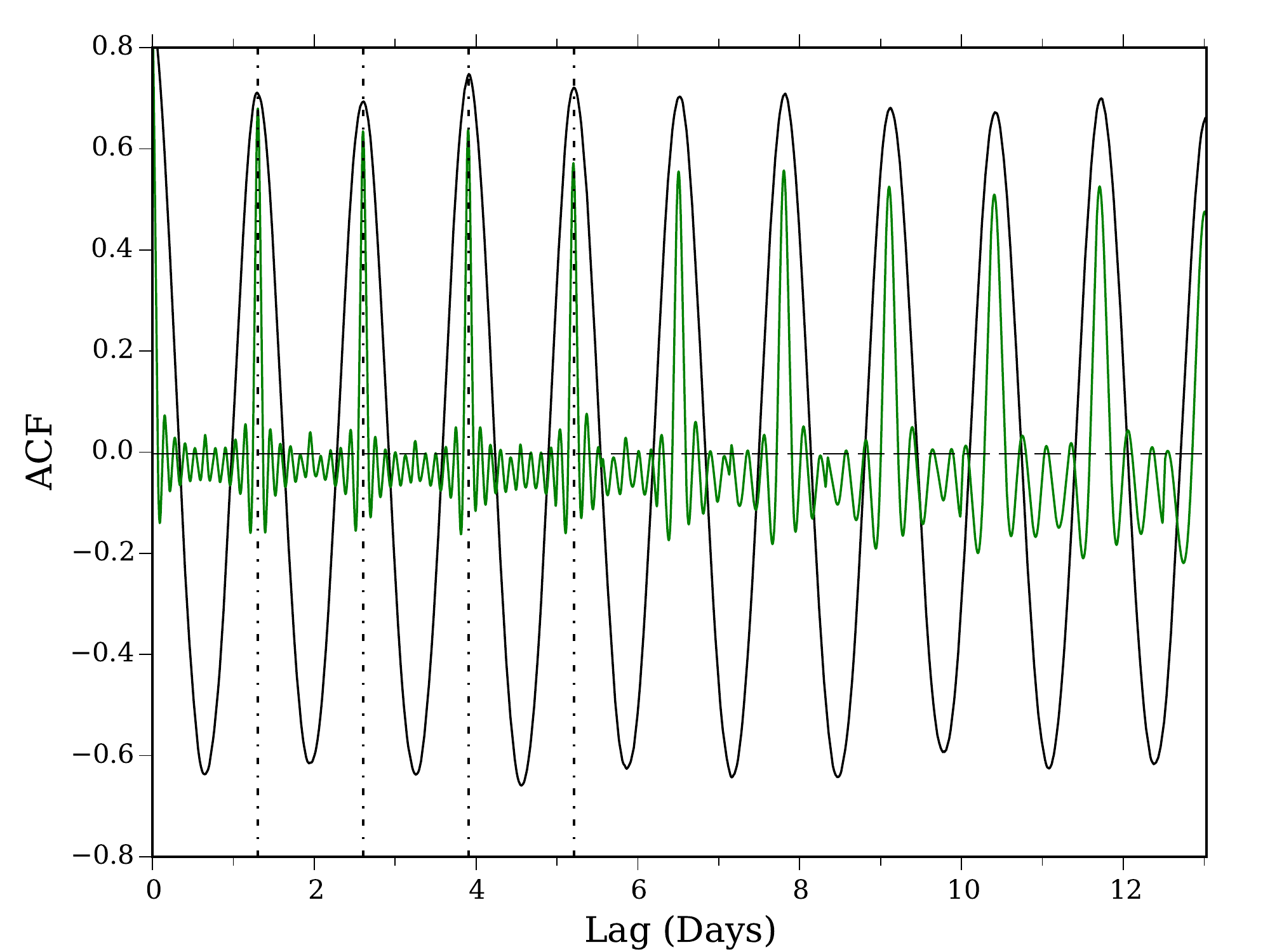}
    \end{subfigure}
    \caption{Autocorrelation functions for \targ (left) and \targt (right), where only the systematics from the apparent stellar motion on the CCD and a $30$ day Epanechnikov \citep[][]{hastie2009elements} filter have been removed. Vertical broken lines indicate the first four maxima of a given periodicity. For \targ we have in red added a $1.23$ day Epanechnikov smoothed version to highlight the underlying ${\sim}7$ day periodicity. In green we shown the so-called narrowed autocorrelation (NACF) where the the response at a given lag is formed from the mean of 10 equally spaced lags of the ACF \citep[][]{nacf_ref,nacf_ref2}. The narrow peaks in the NACF is a testament to the strong regularity in and stability of the periodic signals.}
\label{fig:acf}
\end{figure*}
We were unable to obtain a good estimate of the stellar rotation from the rotational splittings of mode frequencies. The surface rotation could, however, be studied using K2 light curves, which show signs of spot modulation. We estimated the surface rotation period in three complimentary ways \citep[][]{2015MNRAS.450.3211A}, namely, from the imprint in the power spectrum at low frequencies \citep[see, \eg,][]{2013A&A...557L..10N}, from the autocorrelation function (ACF) \citep[see, \eg,][]{2013MNRAS.432.1203M}, and from a Morlet wavelet analysis of the light curve \citep[see][]{2010A&A...511A..46M,2014A&A...572A..34G,2016MNRAS.456..119C}. All three methods agree for both stars.
In \fref{fig:acf} we show the ACFs of the light curves where only the correction from the apparent movement on the CCD has been removed, \ie, any long term trends will be retained. 

For \targ we see the presence of two periodicities (see \fref{fig:acf}), viz., ${\sim}0.8$ and ${\sim}7$ days. Which of these periodicities represent the rotation is somewhat ambiguous, but from the wavelet analysis the ${\sim}0.8$ day period is favoured. In the fast scenario ($P_{\rm rot}\approx 0.8$ days) the star would rotate at about ${\sim}18\%$ of break-up\footnote{computed as $v_{\rm crit}\approx\sqrt{2GM/3R_{\rm p}}$ and assuming that the polar radii $R_{\rm p}$ can be approximated by the non-rotating radius \citep[][]{2009pfer.book.....M}.} and have an inclination of ${\sim}14\pm 2^{\circ}$. Concerning the projected rotation rate used to obtain the inclination \citet[][]{2009A&A...498..949M} gives an independent and corroborating value of $v\sin\, i_{\star}= 20.4 \pm 2.0\, \rm km\ s^{-1}$. The longer period signal might be interpreted as the effect of beating of close frequencies from differential rotation --- indeed, signal is seen in the power spectrum in a group of peaks at ${\sim}13.8\, \rm \mu Hz$ (${\sim}0.84$ days), at ${\sim}28.8\, \rm \mu Hz$ (close to sum of grouped peaks), and at ${\sim}0.6-1.6\, \rm \mu Hz$ (close to differences of grouped peaks) as would be expected for a beating signal \citep[see, \eg,][]{2014A&A...562A.124M}. 
Removing the ${\sim}7$ day signal from the ACF we obtain a spot decay time of ${\sim}10$ days \citep[see][their Figure 4]{2013MNRAS.432.1203M}.
In terms of $(B-V)_0$, or mass, the fast scenario is supported by the proximity to the Kraft break \citep[][]{1967ApJ...150..551K}, which marks where the surface convection zone becomes too shallow to produce a significant braking through a magnetised wind and the observed rotation rate thus becomes highly dependent on the initial rotation rate \citep[see, \eg,][]{2013ApJ...776...67V}. 
Given that we see solar-like oscillations and possible signals from spots the star must have a convective envelope, but given the relatively young age of the star (compared to its MS lifetime) it is conceivable that spin-down has not had time to take effect. The slow scenario of $P_{\rm rot}\approx7$ days matches the rotation period of $P_{\rm rot}\approx 6$ days predicted from the $P_{\rm rot} - (J-K_s)$ relation for the Hyades by \citet[][]{2011MNRAS.413.2218D}, but we note that the $(J-K_s)$ colour of our target is at the limit of the calibration for this relation. In terms of other gyrochronology relations from, for instance, \citet[][]{2007ApJ...669.1167B} and \citet[][]{2008ApJ...687.1264M} the fast scenario is supported.

If the rotation rate follows the fast scenario higher order effects should perturb the oscillation frequencies \citep[see, \eg,][]{1998ESASP.418..385K,2006A&A...455..621R,2012A&A...542A..99O}. \citet[][]{0004-637X-721-1-537} describe the effect on the ridges of the \'{e}chelle diagram from including rotation in a perturbative manner and near-degeneracy effects, and find among other effects a shift in the $\rm\delta_{01}$ spacing between ridges. As mentioned in \sref{sec:pb} we find $\rm\delta_{01} \approx 0\, \mu Hz$ from the ridge averages shown in \fref{fig:echelle}; from a range of models matching the star in terms of mass, age, and metallicity we derive a median $\rm\delta_{01} \approx 2.3 \pm 0.6\, \mu Hz$, where the uncertainty is given by the median-absolute-deviation of the individual model values. This difference could potentially be caused by rotation, but an in-depth analysis of such higher order effects is beyond the scope of the current paper.

For \targt only a single period of ${\sim}1.28$ days is seen in the ACF (right panel of \fref{fig:acf}), which corresponds to ${\sim}12\%$ of break-up. Comparing the measured $v\sin\, i_{\star}$ to the estimated rotational velocity suggests that the star is seen at an angle of $i_{\star}\approx 90^{\circ}$, that is, edge-on. Concerning the projected rotation rate used to obtain the inclination \citet[][]{1963ApJ...137..301G}, \citet[][]{1965ApJ...142..681K}, and \citet[][]{1982bscf.book.....H} \citep[see][]{1995yCat.5050....0H}, give independent and largely corroborating values of $v\sin\, i_{\star}= 50\, , 55\,$, and $53\, \rm km\ s^{-1}$.
\targt was studied by \citet[][]{1995MNRAS.277.1404K} in a search for $\gamma$-Doradus Type variables in the Hyades, where the authors postulate that the detected variability is likely due to spot modulation\footnote{according to the SIMBAD database this study is the reason why the star is listed in \citet[][]{2009yCat....102025S} (and hence SIMBAD) as an ellipsoidal variable star, which it is not.}. Curiously, these authors find a periodicity of $1.4336$ days, albeit from only 76 data points over a 20 day period.

An additional assessment of the stellar activity signal comes with the measured coronal activity in terms of X-ray luminosity. From the \textit{ROSAT} X-ray hardness ratio measurements in the $0.1-2.4$ keV band by \citet[][]{1999A&A...349..389V} we obtained for \targ an X-ray to bolometric luminosity of $\log_{10} R_X=-4.82\pm 0.08$, with $R_X=L_X/L_{\rm bol}$. Here we used the conversion between \textit{ROSAT} counts and hardness ratio to flux by \citet[][]{1995ApJS...99..701F} and \citet[][]{1995ApJ...450..392S}, and the luminosity estimated in \sref{sec:atmos}.
The above value corresponds largely to those from the $0.2-2.8$ keV band \textit{Einstein Observatory} and \textit{ROSAT} All-Sky Survey (RASS) measurements by, respectively, \citet[][]{1981ApJ...249..647S} and \citet[][]{1995ApJ...448..683S} after converting when appropriate to the \textit{ROSAT} $0.1-2.4$ keV band using PIMMS\footnote{The Chandra Portable Interactive Multi-Mission Simulator, \url{www.cxc.harvard.edu/toolkit/pimms.jsp}}.
For \targt we derive from measurements by \citet[][]{1999A&A...349..389V} a value of $\log_{10} R_X \approx -5.47 \pm 0.16$; Coronal X-ray measurements from the RASS by \citet[][]{1998A&AS..132..155H} and \textit{ROSAT} measurements from \citet[][]{1995ApJ...448..683S} largely agree with this estimate. 
\citet[][]{2011ApJ...743...48W} offers a relation between $R_X$ and the Rossby $Ro$ number \citep[see also][]{2003A&A...397..147P,2014ApJ...795..161D}, where $Ro$ is defined as $P_{\rm rot}/\tau_c$ with $\tau_c$ being the mass-dependent convective turnover time-scale. In the following we use the $\tau_c(M)$ relation from \citet[][]{2011ApJ...743...48W} to determine $Ro$, with the mass from the seismic modelling (\sref{sec:model}). 

For \targ the two different scenarios for the rotation rate corresponds to Rossby numbers of $Ro\sim 0.10\pm0.01$ (fast) and $Ro\sim 0.83\pm0.06$ (slow). From \citet[][]{2011ApJ...743...48W} one should for $Ro\sim 0.83$ expect a level of $R_X\approx -5.18\pm0.24$, and for $Ro\sim 0.10$ the star should fall in the saturated regime with $\log_{10}R_X\approx -3.13\pm0.08$. For \targt one would expect a value of $\log_{10}R_X \approx -3.38\pm0.28$.
As seen the measured levels disagree with those expected for \targt and the fast scenario for \targ. The $(B-V)$ colours of the stars, with values of $(B-V)=0.42$ (\targ) and $(B-V)=0.41$ (\targt), do, however, also place the stars outside the calibration range adopted by \citet[][]{2011ApJ...743...48W}. Comparing instead to \citet[][]{1987ApJ...321..958V} who include hotter stars we find that the two stars conform with an expected range of $R_X\approx -4.5\, {\rm to} -5.5$. The relatively low levels of chromospheric activity also agrees with the results of \citet[][]{1991ApJ...380..200S} and \citet[][]{1993A&A...269..446S}, who both include \targt in their analysis. These authors find that activity is reduced for stars earlier than ${\sim}$F5, likely due to an inefficient dynamo operating in the shallow convection zone of such early-type stars. In a study of F5-type stars in the Hyades \citet[][]{2002ApJ...569..941B} finds that $(B-V)\approx 0.42-0.43$ marks a transition region in the dependence of X-ray flux with $\vsini$ (with a decreasing X-ray flux with increasing \vsini), and in the onset of an efficient magnetic braking. Both stars thus seem to be in a very interesting region in terms of rotation and activity.      

For both stars we further assessed the mean activity level using the activity proxy $\langle \rm S_{ph,{\it k}=5}\rangle$ as defined in \citet[][]{2010Sci...329.1032G} and \citet{2014A&A...562A.124M,2014JSWSC...4A..15M}. For \targ we adopted the fast scenario for the period used in calculating the activity proxy. We obtained values of $\langle \rm S_{ph,{\it k}=5}\rangle = 273 \pm 6$ ppm (\targ) and $\langle \rm S_{ph,{\it k}=5}\rangle = 249 \pm 7$ ppm (\targt).
Comparing these with \citet[][their Figure~10]{2014A&A...572A..34G} it is clear that the two stars occupy a region of the $P_{\rm rot}$--$\langle \rm S_{ph,{\it k}=5}\rangle$ space that is unexplored with data from the nominal mission --- this likely stems from the sparsity in the number of young, hot, stars that were suggested for observations for the sake of detecting solar-like oscillations.


\subsection{Asteroseismic modelling}
\label{sec:model}

The two targets analysed here only provide us with limited seismic information, that is, only the average seismic parameters $\Delta\nu$ and $\nu_{\rm max}$. These were used together with estimates of the two stars' metallicity and effective temperatures to determine the global parameters of the stars using grid based searches.
Three pipelines were used in the modelling --- the Yale-Birmingham pipeline  \citep[YB; ][]{2010ApJ...710.1596B, 2012ApJ...746...76B,2011ApJ...730...63G}, the Bellaterra Stellar Parameters Pipeline \citep[BeSPP;][Serenelli (in prep.)]{2013MNRAS.429.3645S}, and the Bayesian Stellar Algorithm pipeline \citep[BASTA; ][]{2015MNRAS.452.2127S}. 
Three different grids of models were used in the case of YB, with models from the Dartmouth group \citep[][]{2008ApJS..178...89D}, the Yonsei-Yale ($Y^2$) isochrones \citep[][]{2004ApJS..155..667D}, and the Yale Stellar Evolution Code \citep[YREC;][]{2008Ap&SS.316...31D} as described by \citet[][]{2012ApJ...746...76B} (YREC2). In all cases \dnu for the YB models were calculated using the simple scaling relation between \dnu and density (\ie, $\dnu\propto\sqrt{M/R^3}$). BeSPP and BASTA used grids of models calculated using the Garching Stellar Evolution Code \citep[GARSTEC;][]{2008Ap&SS.316...99W}. For BeSPP and BASTA model values for $\Delta\nu$ were calculated using both the scaling relation and individual frequencies of radial modes. In cases where the \dnu scaling relation was used, the corrections given in \citet[][]{2011ApJ...742L...3W} (for YB) and Serenelli et al. (2016, in prep.) (for BeSPP and BASTA) were applied to correct for the deviations of $\Delta\nu$ values from the usual scaling relations. The value of \numax was computed using the usual scaling scaling relation ($\numax\propto g/\sqrt{\teff}$). 
Further details of the pipelines, grids, and scaling relations are described in \citet[][]{2014ApJS..210....1C} and \citet{2015MNRAS.452.2127S}.

From the grid-based modelling (GBM) we obtain for \targ values of $M=1.41\pm 0.06\, \rm M_{\odot}$, $R= 1.48\pm 0.03\, \rm R_{\odot}$, $\rho= 0.60\pm  0.15 \,\rm g\, cm^{-3}$, and $t = 1020\pm 387\,\rm Myr$; for \targt we obtain $M=1.47\pm 0.06\, \rm M_{\odot}$, $R= 1.61\pm 0.03\, \rm R_{\odot}$, $\rho= 0.50 \pm 0.13\,\rm g\, cm^{-3}$, and $t = 1132\pm 304\,\rm Myr$. The reported values are those obtained from the BASTA pipeline using the SPC \teff and \feh. We have added in quadrature to the formal uncertainties a systematic uncertainty given by the root-mean-square difference between the reported BASTA values and those obtained from the other pipelines and spectroscopic inputs.

Both grid-based age estimates are seen to be slightly higher than
those normally derived from isochrone fitting to the full
colour-magnitude diagram (CMD). This difference is not completely
unexpected, and has its origins in the limited nature of the data
available to us here, as we now go on to explain. Nevertheless, as we
shall also see, including the asteroseismic parameters for these
main-sequence stars gives much better constraints on the fundamental
properties than would be possible from CMD fitting of main-sequence
stars alone.

\begin{figure*}
    \centering
    \begin{subfigure}
        \centering
        \includegraphics[width=\columnwidth]{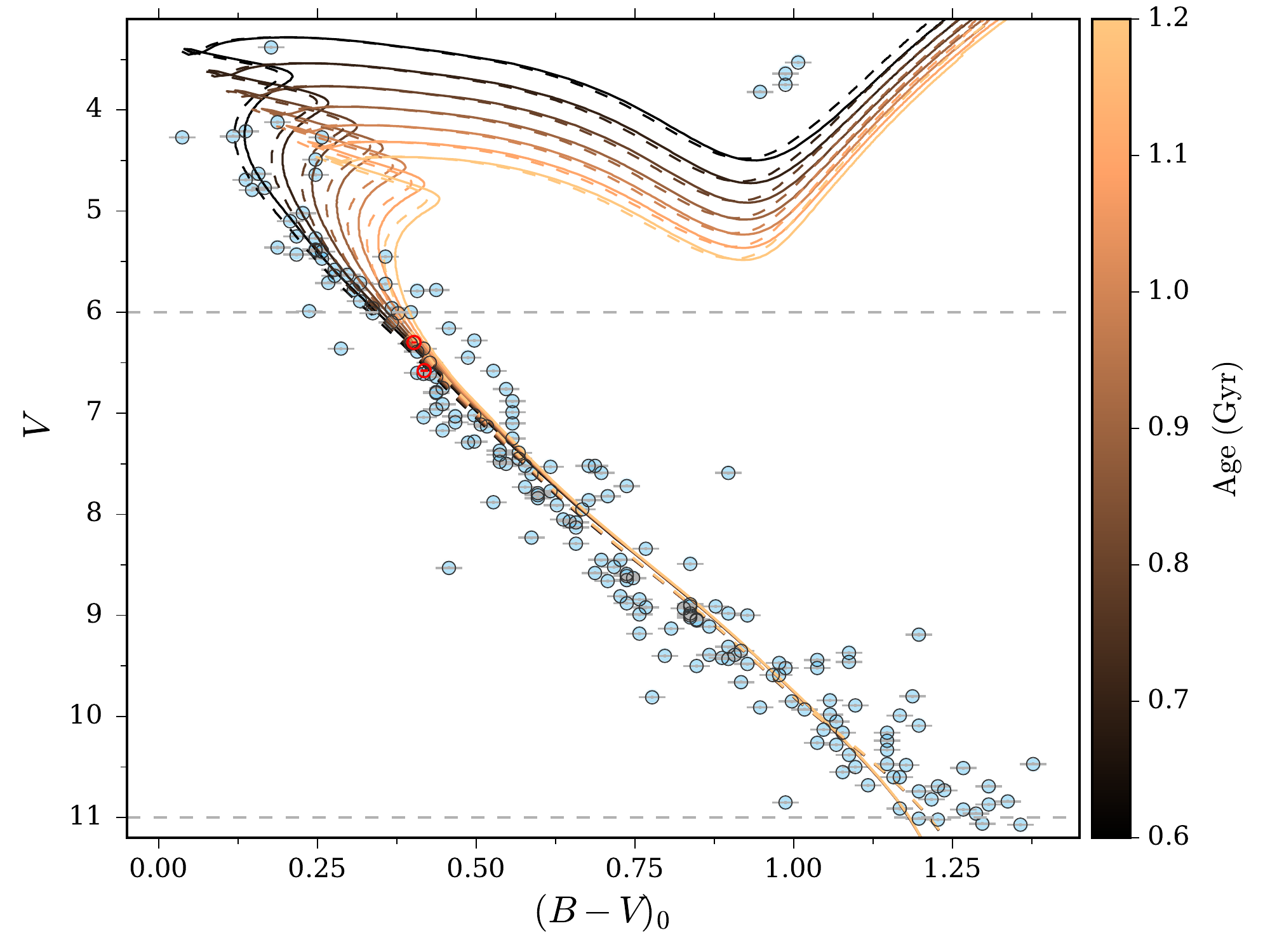}
    \end{subfigure}%
    \hfill
    \begin{subfigure}
        \centering
        \includegraphics[width=\columnwidth]{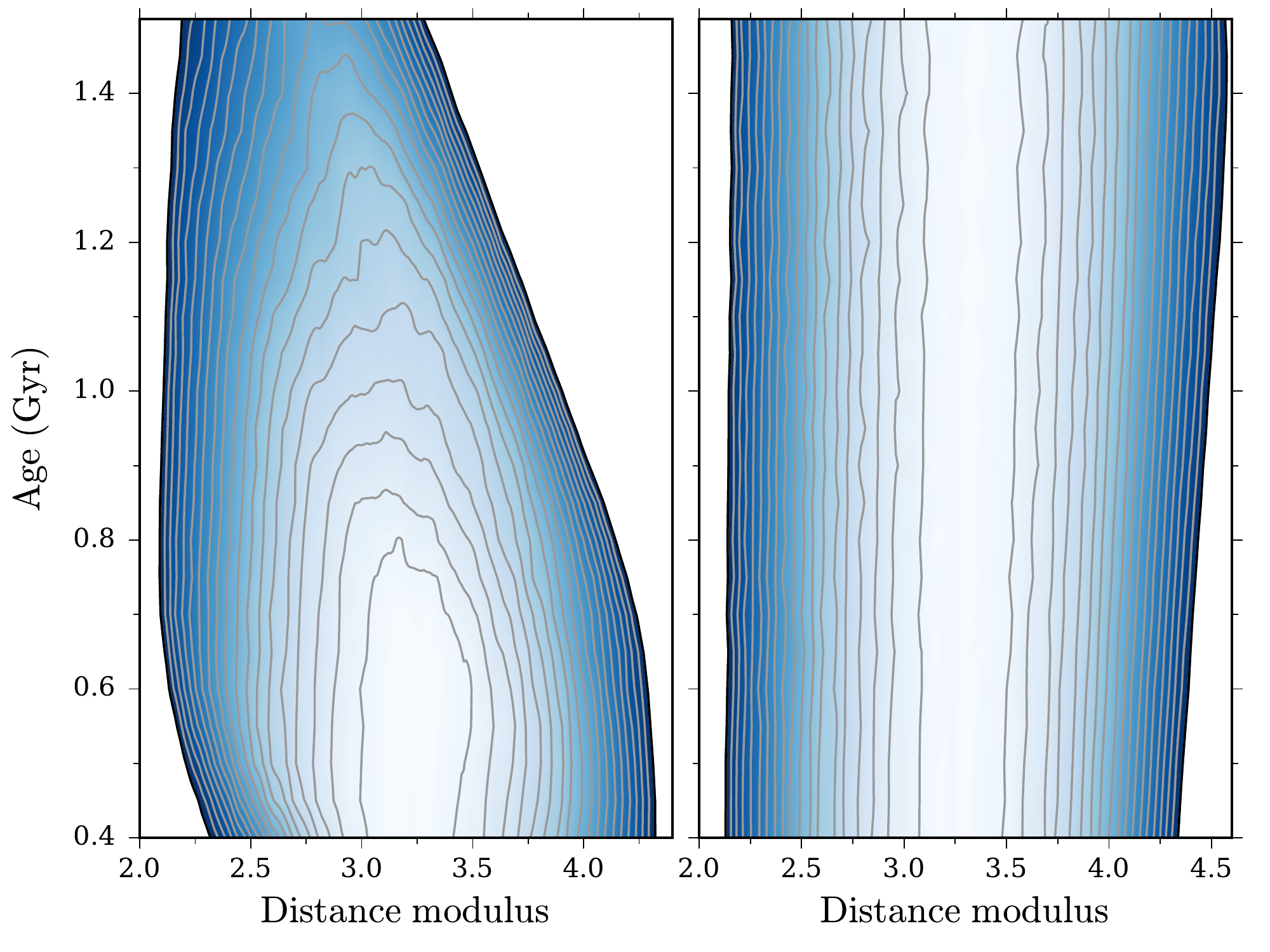}
    \end{subfigure}

    \caption{Left: colour-magnitude diagram (CMD) of Hyades stars,
      with parameters adopted from
      \citet[][]{1995ApJ...448..683S}. The colour of the GARSTEC
      plotted isochrones indicate to age; full-line isochrones have
      $\feh=0.2$, while dashed-lined ones have $\feh=0.15$. For all
      isochrones we adopted for this plot a distance modulus of
      $(m-M)=3.25$. We adopted $E(B-V)=0.003\pm 0.002$
      \citep[][]{1980AJ.....85..242T} and $R_V\equiv A_V/E(B-V) = 3.1$
      to de-redden the values from
      \citet[][]{1995ApJ...448..683S}. The two stars analysed in this
      work are given by the red markers, here using updated colours
      from \citet[][]{2006AJ....132..111J}. Right: $\chi^2$ surfaces
      for CMD fits to stars with $V<11$ (left panel) and $6<V<11$
      (right panel) as a function of age and distance modulus, going
      from high $\chi^2$-values in dark blue to low values in light
      blue. For both cases shown $\feh=0.2$.}

\label{fig:cmd_fit}
\end{figure*}

We begin by recalling that CMD fitting of clusters works well only
when data are available that span a range of evolutionary states, i.e,
including turn-off stars and also red-giant-branch stars.  The
left-hand panel of \fref{fig:cmd_fit} shows the CMD of the Hyades
using data from \citet[][]{1995ApJ...448..683S}. The plotted
isochrones are from GARSTEC, with colour indicating age (see the
sidebar) and linestyle indicating metallicity (full-line isochrones
have $\feh=0.2$, while dashed-line isochrones have $\feh=0.15$).  The
distance modulus adopted for this plot (and subsequent analysis) was
$(m-M)=3.25$. We used $E(B-V)=0.003\pm 0.002$
\citep[][]{1980AJ.....85..242T} and $R_V\equiv A_V/E(B-V) = 3.1$ to
de-redden the values from \citet[][]{1995ApJ...448..683S}. Note that
the two stars with asteroseismic detections are plotted in red (here
using updated colours from \citet[][]{2006AJ....132..111J}).

We used the BeSPP pipeline to fit the observed CMD data in
\fref{fig:cmd_fit} to the GARSTEC isochrones.  The right-hand panels
show the resulting $\chi^2$ surfaces for two fits: one where we
limited data to the main-sequence only ($6<V<11$, right) and
another where we used all the available data (stars with $V<11$, left). For
both cases shown we adopted $\feh=0.2$. We see that limiting to the
main-sequence only provides no discernible constraint on age. The
constraints are of course even weaker if we perform CMD fits using the
two asteroseismic stars only (again with no seismic data). In
contrast, we obtain good constraints on the age, and optimal values
that agree with the canonical literature values, when we include
targets beyond the main-sequence (see also
\citealt[][]{1998A&A...331...81P} and
\citealt[][]{2004ApJ...600..946P}).  Unfortunately, stars close to the
turn off are likely to be too hot to show solar-like
oscillations. Nevertheless, we see that the asteroseismic results
obtained on the two stars -- albeit using average asteroseismic
parameters only -- give much better constraints than those provided by
the CMD fits to non-seismic data on main-sequence stars
alone.

That the age constraints from the asteroseismic results are not tighter still reflects the nature of the average asteroseismic parameters. Both depend (in whole or large part) on different combinations of ratios of mass and radius --- they thus lack explicit information on core properties and this has an impact on age estimates for stars in the relatively slow MS phase \citep[][]{2011ApJ...730...63G,2014ApJS..210....1C}. Much tighter constraints are possible on the low-age part of the MS when using individual oscillation frequencies \citep[][]{2015MNRAS.452.2127S}.

Nevertheless, we still see a bias in the asteroseismic age estimates, and some of this arises from the way in which ages are estimated using a probabilistic approach when matching to isochrones (or grids) of stellar models. If a star lies equally close to two isochrones in terms of input parameters the most likely will be chosen based on evolutionary speed. Therefore, without prior knowledge, one is much more likely to find a star that belongs to an older (say over 1\,Gyr)
isochrone because evolution is slower than for a sub 1-Gyr isochrone. Two of our pipelines (BASTA and BeSPP) use Bayesian schemes when computing the posterior parameter distributions, and here correct for the density of points in the adopted grids to make a proper marginalisation --- this correction explicitly introduces the effect of evolutionary speed (see, \eg, \citet[][]{2004MNRAS.351..487P} and \citet[][]{2005A&A...436..127J} for examples and further discussion). 

There are two main reasons for the bias, one easy to remove and one more fundamental. The first reason is that at low ages, the distribution function of ages for a given star cuts off abruptly at zero, biasing the result to higher ages. This effect can be mitigated to some extent by using the logarithm of the ages, but this does not remove the bias completely. The second reason for the bias is more fundamental, and has to do with the fact that on the main-sequence, stars within a small metallicity range can have many different ages for a given range of temperature and luminosity (or in the asteroseismic context a given range of $\nu_{\rm max}$). In other words, isochrones of many different ages can pass through the error-box. As described above, evolutionary speed makes it much more likely to encounter an older than a younger star, and therefore the results of any grid-based modelling will have a fundamental bias towards higher ages if no prior on age is adopted. Figure~B1 shows this clearly. The bias can be reduced if effective temperature and metallicity can be measured to a much better precision. In the case of clusters, having data on more stars in different evolutionary phases of course helps greatly because we can apply the condition that all stars must have the same age, which is essentially the assumption made in determining ages by fitting isochrones to cluster colour-magnitude diagrams. There are also other factors to note. The two stars here are different to many of the stars analysed for asteroseismology in \kp, in terms of being relatively hot, massive, young, and rapidly rotating. This of course raises the question of whether assumptions made regarding the mapping of the average asteroseismic parameters to stellar properties are incorrect for these stars? The results suggest there is not a significant bias.  First, the relationship of \dnu to \numax follows that shown by the asteroseismic cohort of \kp stars.  We also examined the potential impact of rotational mixing on \dnu, which is unaccounted for in the models we used, by looking at differences in stellar MESA models \citep[][]{2011ApJS..192....3P} with and without
convective core overshoot. We found no appreciable change in \dnu from varying the amount of overshoot, which conforms with the results reported by \citet[][]{2010A&A...519A.116E} who studied the effect of adding rotational mixing to a $\rm 1M_{\odot}$ model. We therefore adopt the assumption that the model values of \dnu and \numax are representative of what would be found for slowly rotating stars. We also remind the reader that in \sref{sec:pb} it was found that rotation should not affect our ability to extract a good estimate of \dnu.

The above of course also goes to the issue of the physics used in our stellar models. Might missing physics be a cause of the bias? The obvious question we can answer in relation to this is whether, when we use our adopted models, we are able to recover the canonical age estimate when presented with the usual observational CMD data as input (\ie, colours and an assumed distance modulus and metallicity as input). As discussed above (\fref{fig:cmd_fit}), we have demonstrated that when BeSPP is coupled to GARSTEC, we recover a satisfactory age. That does not of course say that the physics is indeed correct.

Recently, \citet[][]{2015ApJ...807...24B,2015ApJ...807...25B,2015ApJ...807...58B} performed an isochrone analysis which included rotation via the models of \citet[][]{2012A&A...537A.146E} and \citet[][]{2013A&A...553A..24G}. By adding rotation, which in the adopted models had the effect of lengthening the MS lifetime, these authors find a slightly higher age than the consensus, namely, $t\sim 750\pm 100$ Myr. This result rests on the same handful of upper MS ($M>1.7\rm M_{\odot}$) turn-off stars that guided the isochrone fittings by \citet[][]{1998A&A...331...81P}.


\subsection{Distances}
\label{sec:dist}

\begin{figure}
\centering
\includegraphics[width=\columnwidth]{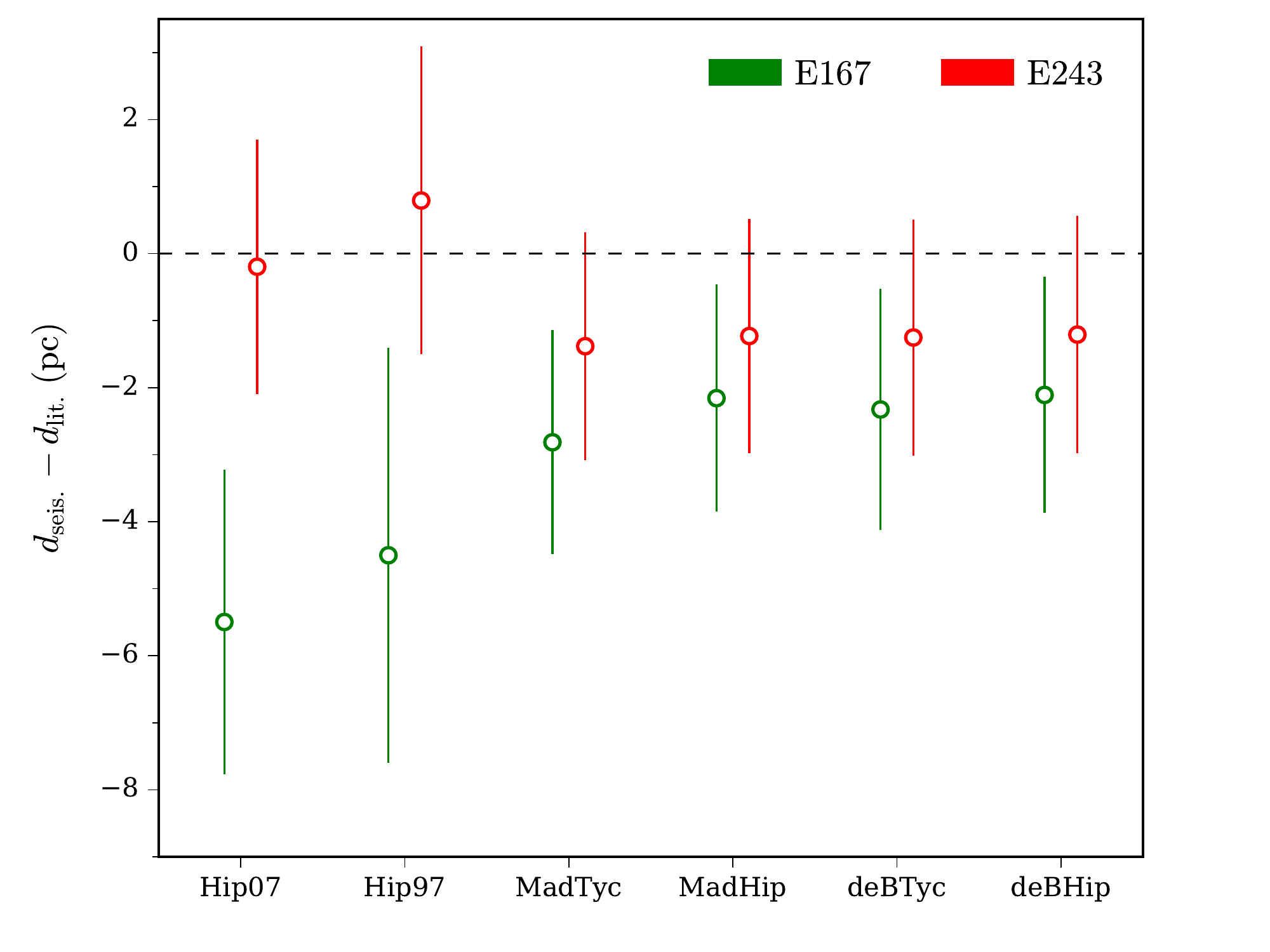}
\caption{Comparison of distances obtained from the seismic radii (average from different spectroscopic inputs and pipelines) and the IRFM angular diameter with those determined from parallaxes in the literature.}
\label{fig:dist}
\end{figure}

With the seismic solution for the stellar radii and an angular diameter from the IRFM, we can estimate the seismic distance to the cluster as follows: 
\begin{equation}\label{eq:sd}
d_{\rm seis} = C \frac{2R_{\rm seis}}{\theta_{\rm IRFM}}\, ,
\end{equation} 
where $C$ is a conversion factor to parsec \citep[see][]{2012ApJ...757...99S,2014MNRAS.445.2758R}.
We find seismic distances of $d_{\rm seis}=46.9\pm 1.5$ pc (\targ) and $d_{\rm seis}=45.2\pm1.6$ pc (\targt).
In \fref{fig:dist} we compare these to the distances derived from trigonometric parallaxes from \textit{Hipparcos} by \citet[][]{2007A&A...474..653V} (Hip07), \citet[][]{1997A&A...323L..61V} (Hip97), those from \citet[][]{2001A&A...367..111D} using secular parallaxes from Tycho-2 \citep[][]{2000A&A...355L..27H} (deBTyc) or \textit{Hipparcos} \citep[][]{1997A&A...323L..61V} (deBHip), and those from \citet[][]{2002A&A...381..446M} using secular parallaxes as above (MadTyc/MadHip).
We find that all parallax distances for \targt match the seismic ones reasonably well; for \targ all distances are ${>}1\sigma$ larger than the seismic ones. 


\section{Discussion and outlook}
\label{sec:dis}

We have presented the asteroseismic results on two cool main-sequence
stars belonging to the Hyades open cluster. These are the first ever
detections of solar-like oscillations in main-sequence stars in an
open cluster. Both stars are very likely fast rotators ($P_{\rm
  rot}<2$ days), marking them out as quite different from the older,
more slowly rotating cool main-sequence stars that dominated the
asteroseismic cohort from the nominal \kp\ mission.

\begin{figure}
\centering 
\includegraphics[width=\columnwidth]{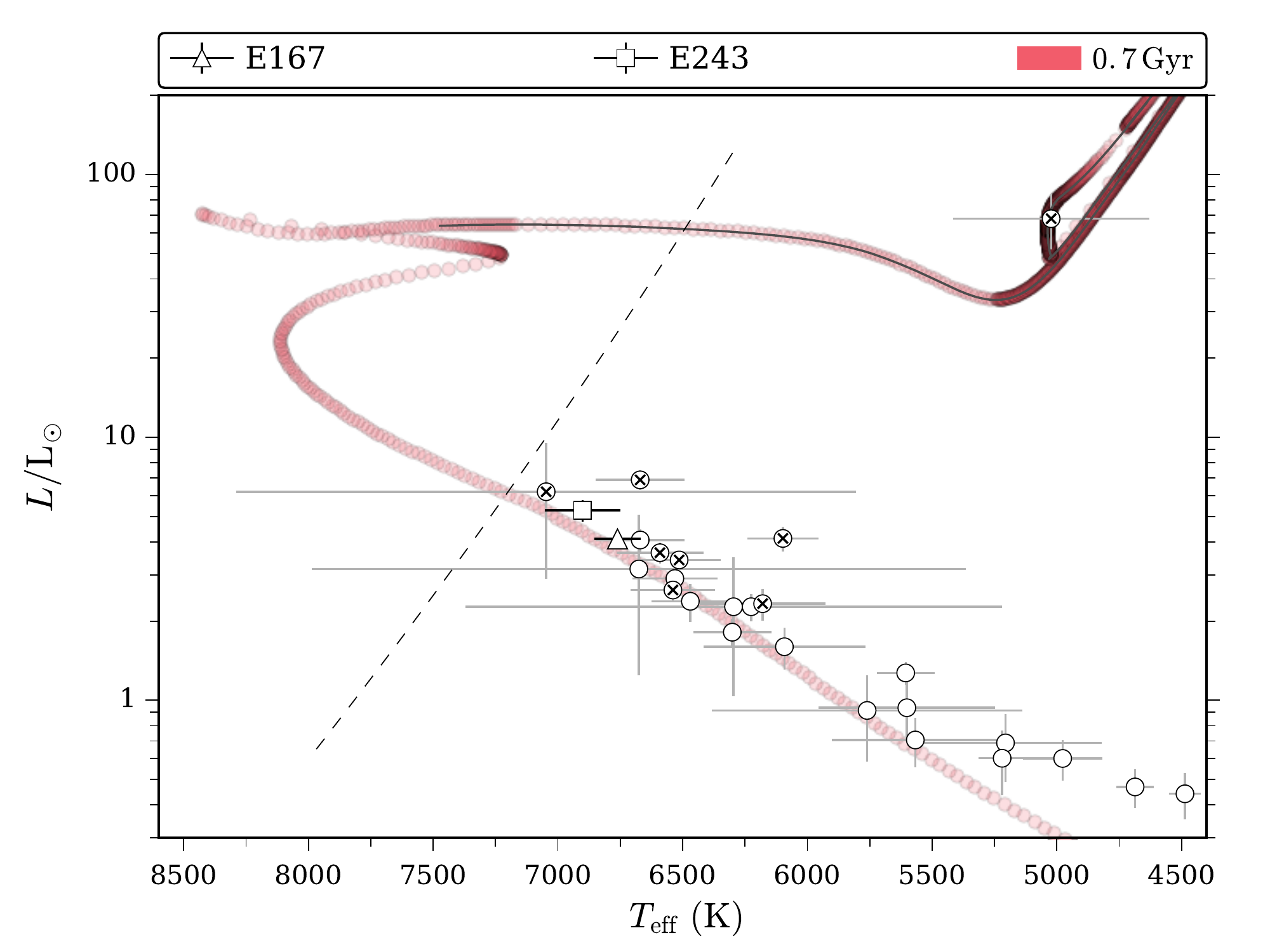}
\vspace{-0.5cm} 
\caption{BaSTI isochrone \citep[][]{2004ApJ...612..168P} with an age of
  700 Myr. For this isochrone standard BaSTI input
  physics was adopted, but with overshoot on the MS and with adopted heavy and
  Helium mass fractions of $Z=0.0198$ and $Y=0.273$. The triangle and
  square symbols indicate our two targets (see legend), with \teff
  given by the SPC and with the difference to the IRFM value added in
  quadrature to the uncertainty. The luminosities for the two targets
  are given by these \teff values and the seismic radii. Circular
  markers indicate stars that could be observed in Campaign 13 for
  which we predict detectable solar-like oscillations. The stars that
  are further marked with a cross will be amenable to interferometric
  observations with PAVO@CHARA
  \citep[][]{2005ApJ...628..453T,2008SPIE.7013E..24I}. The dashed line
  gives the red edge of the classical instability strip, as defined by
  \citet[][]{2000ASPC..210..215P}. The parts of the isochrone
  overlaid with a solid black line indicate where \numax values
  computed from the standard scaling by
  \citet[][]{1995A&A...293...87K} drops below the long-cadence Nyquist
  frequency of ${\sim}283\, \rm\mu Hz$.}

\label{fig:iso} 
\end{figure}

The K2 mission is scheduled to re-observe the Hyades cluster in C13,
providing an unprecedented opportunity to expand the asteroseismic
cohort, potentially to stars for which we can do detailed modelling on
individual frequncies (something that is very challenging for the two
stars reported here). We have indicated in \fref{fig:iso} the stars
from C13 for which we predict a detection of solar-like oscillations
(including predicted marginal detections). The estimates of $L$ used
here were computed from \textit{Hipparcos} parallaxes
\citep[][]{2007A&A...474..653V}, while \teff values were computed from
the colour-\teff relations of \citet[][]{2010A&A...512A..54C}.

Unfortunately neither of the targets analysed in this paper is
predicted to lie on active silicon in C13\footnote{using the
  \texttt{K2FOV} tool;
  \url{www.keplerscience.arc.nasa.gov/software.html}}.  We find,
however, that $55$ of the Hyades members from
\citet[][]{1998A&A...331...81P} will be on silicon in C13; of these we
estimate ${\sim}22$ will have \kp magnitudes in the range $\rm Kp =
6-9.5$, $T_{\rm eff}\lesssim 6300$ K, and rotational periods in the
range $P_{\rm rot} \approx 6-15$ days. Based on knowledge of K2 noise
properties asteroseismic analysis of these targets should be feasible
(see \citealt[][]{2015ApJ...809L...3S}; \citealt{2015arXiv151206162V};
Lund et al., in press). A joint analysis may provide constraints on
the cluster age, especially if individual frequencies or even just an
estimate of the core-sensitive small frequency separation
$\delta\nu_{02}$ can be obtained in some stars \citep[][]{1993ASPC...42..347C}.
Moreover, for several stars independent constraints may be obtained
from interferometry with PAVO@CHARA
\citep[][]{2005ApJ...628..453T,2008SPIE.7013E..24I}.  One of the C13
targets is a giant and can comfortably be observed in long-cadence
(LC; $\Delta t\approx 30$ min) --- this star (HIP 20885; 77 Tau) would
be valuable to constrain the cluster age, especially if combined with
the MS targets and the two C4 giants analysed by White et al. (in
prep.).

\section*{Acknowledgments}
\footnotesize

We acknowledge the dedicated team behind the \kp and K2 missions, without whom this work would not have been possible.
We thank Daniel Huber and Benoit Mosser for useful comments on an earlier version of the paper.
M.N.L. acknowledges the support of The Danish Council for Independent Research | Natural Science (Grant DFF-4181-00415).
M.N.L. was partly supported by the European Community's Seventh Framework Programme (FP7/2007-2013) under grant agreement no. 312844 (SPACEINN), which is gratefully acknowledged.
Funding for the Stellar Astrophysics Centre (SAC) is provided by The Danish National Research Foundation (Grant DNRF106). The research was supported by the ASTERISK project (ASTERoseismic Investigations with SONG and \kp) funded by the European Research Council (Grant agreement no.: 267864).
W.J.C., G.R.D, and A.M. acknowledge the support of the UK Science and Technology Facilities Council (STFC). 
S.B. acknowledges partial support from NASA grant NNX13AE70G and NSF grant AST-1514676.
A.M.S. is partially supported by grants ESP2014-56003-R and ESP2015-66134-R (MINECO).
V.S.A. acknowledges support from VILLUM FONDEN (research grant 10118).
R.A.G. acknowledges the support from the ANR program IDEE (n. ANR-12-BS05-0008) and the CNES.
D.W.L acknowledges partial support from the \kp mission under Cooperative Agreement NNX13AB58B with the Smithsonian Astrophysical Observatory.
This research has made use of the Washington Double Star Catalog maintained at the U.S. Naval Observatory; the WEBDA database, operated at the Department of Theoretical Physics and Astrophysics of the Masaryk University; and the SIMBAD database, operated at CDS, Strasbourg, France.


\small
\bibliography{MasterBIB}


\appendix

\section{Radial velocities from CfA}
\label{sec:rv}

\begin{figure*}
\centering
\includegraphics[width=\textwidth]{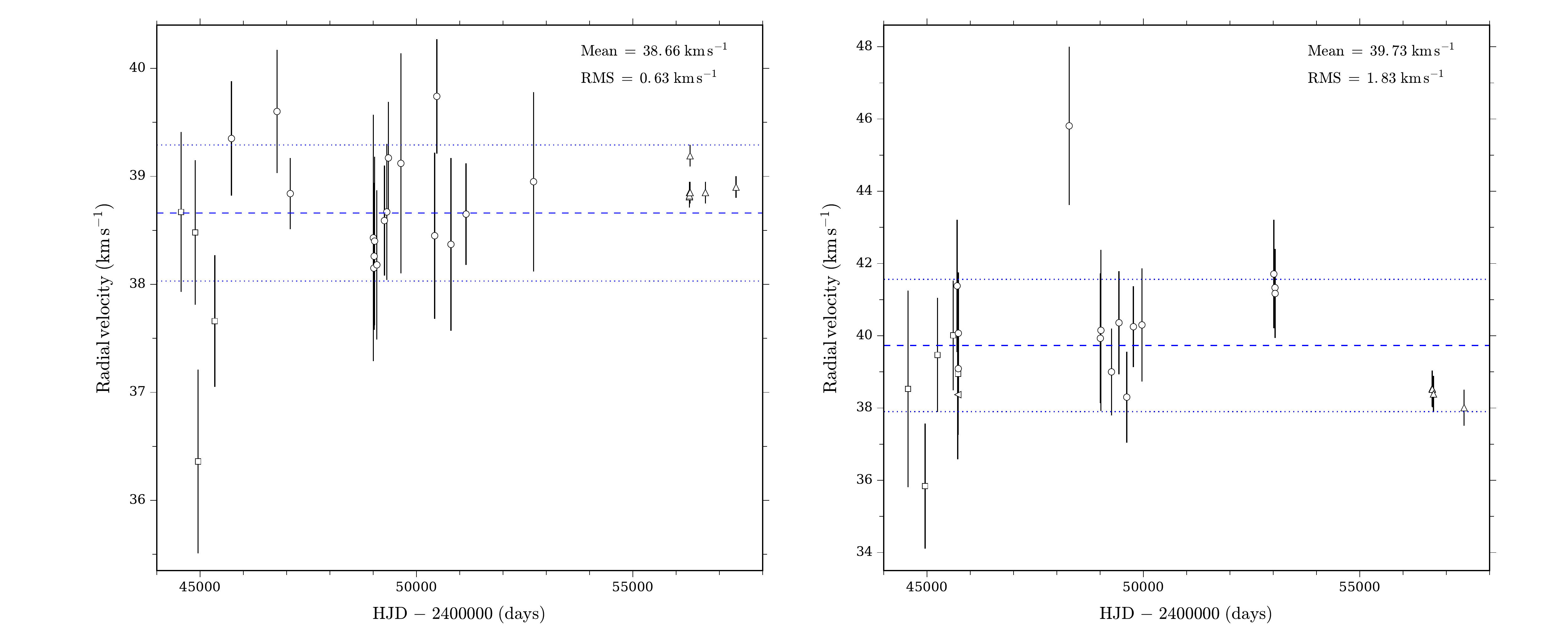}
\caption{Radial velocity data for \targ (left) and \targt (right) from CfA spanning a period of ${\sim}35$ years, see \tref{tab:rv} for the individual data values. The dashed and dotted lines indicate the mean and \textsc{rms} values of the velocities, with the values given in the plots. The markers indicate the different instruments used in obtaining the data, specifically, the MMT Digital Speedometer ($\lhd$); the Tillinghast Reflector Digital Speedometer  ($\square$); the Wyeth Reflector Digital Speedometer ($\circ$); the Tillinghast Reflector Echelle Spectrograph ($\bigtriangleup$).}
\label{fig:vel}
\end{figure*}

\tref{tab:rv} gives the CfA radial velocities obtained over a period of over 35 years. The dates are given in heliocentric Julian date (minus 2400000), and the
radial velocities are on the native CfA system in $\rm km\ s^{-1}$. To put these velocities onto the IAU system, add $0.14\,\rm km\, s^{-1}$. Concerning the estimated internal errors, these are for the CfA Digital Speedometers estimated from the anti-symmetric noise in the correlation function as described in \citet{1979AJ.....84.1511T}; for TRES it is an educated guess based on extensive experience with dozens of hot rapidly rotating stars. 
The velocities are plotted in \fref{fig:vel}. See \sref{sec:data} for additional info on the data.

\begin{table*}
\tabsize
\centering
\caption{\scriptsize Radial velocity data for \targ (28 measurements) and \targt (24 measurements). The four columns give the heliocentric Julian date (minus 2400000), the
radial velocity on the native CfA system in $\rm km\ s^{-1}$, the estimated internal error, and the source of the data. To put these velocities onto the IAU system, add $0.14\,\rm km\, s^{-1}$. For the CfA Digital Speedometers the error is estimated from the anti-symmetric noise in the correlation function as described in the \citet{1979AJ.....84.1511T}; for TRES it is an educated guess based on
extensive experience with dozens of hot rapidly rotating stars.}
\label{tab:rv}
\begin{tabular}{ccccccccccc}
\toprule \\[-0.8em]
\multicolumn{4}{c}{E167} & & & & \multicolumn{4}{c}{E243} \\ 
\cline{1-4} \cline{8-11}\\[-0.5em]
Date & RV & Uncertainty & Source$^{\dagger}$ & & & & Date & RV & Uncertainty & Source$^{\dagger}$\\
(HJD-2400000)  & ($\rm km\ s^{-1}$) & ($\rm km\ s^{-1}$) & &  & & & (HJD-2400000)  & ($\rm km\ s^{-1}$) & ($\rm km\ s^{-1}$) & \\[0.5em]
\cline{1-4} \cline{8-11}\\[-0.2em]
44560.8212 & 38.67 & 0.74 & 2  & & & & 44560.7928 & 38.53 & 2.72 & 2   \\
44887.8537 & 38.48 & 0.67 & 2  & & & & 44954.7683 & 35.84 & 1.73 & 2   \\
44954.8595 & 36.36 & 0.85 & 2  & & & & 45241.9557 & 39.47 & 1.58 & 2   \\
45339.8980 & 37.66 & 0.61 & 2  & & & & 45604.9773 & 40.01 & 1.52 & 2   \\
45725.5179 & 39.35 & 0.53 & 3  & & & & 45694.6675 & 41.38 & 1.83 & 3   \\
46777.6586 & 39.60 & 0.57 & 3  & & & & 45710.5961 & 38.37 & 1.79 & 1   \\
47084.8265 & 38.84 & 0.33 & 3  & & & & 45717.7013 & 38.95 & 1.33 & 2   \\
49004.6987 & 38.43 & 1.14 & 3  & & & & 45721.6767 & 39.09 & 1.83 & 3   \\
49015.5712 & 38.15 & 0.40 & 3  & & & & 45723.5431 & 40.07 & 1.68 & 3   \\
49023.6034 & 38.26 & 0.68 & 3  & & & & 48284.6245 & 45.81 & 2.19 & 3   \\
49033.6385 & 38.40 & 0.78 & 3  & & & & 49004.6898 & 39.93 & 1.80 & 3   \\
49085.5137 & 38.18 & 0.69 & 3  & & & & 49018.5060 & 40.15 & 2.23 & 3   \\
49259.7909 & 38.59 & 0.51 & 3  & & & & 49261.7933 & 39.00 & 1.20 & 3   \\
49314.8078 & 38.67 & 0.63 & 3  & & & & 49435.5377 & 40.36 & 1.43 & 3   \\
49352.6751 & 39.17 & 0.52 & 3  & & & & 49614.9038 & 38.30 & 1.26 & 3   \\
49640.8260 & 39.12 & 1.02 & 3  & & & & 49768.5359 & 40.25 & 1.12 & 3   \\
50421.7596 & 38.45 & 0.77 & 3  & & & & 49965.8878 & 40.30 & 1.57 & 3   \\
50470.5632 & 39.74 & 0.53 & 3  & & & & 53013.6234 & 41.71 & 1.50 & 3   \\
50797.6904 & 38.37 & 0.80 & 3  & & & & 53040.5089 & 41.33 & 1.06 & 3   \\
51146.7703 & 38.65 & 0.47 & 3  & & & & 53043.5878 & 41.17 & 1.23 & 3   \\
52706.5335 & 38.95 & 0.83 & 3  & & & & 56673.7786 & 38.54 & 0.50 & 4   \\
56308.7281 & 38.81 & 0.10 & 4  & & & & 56675.8239 & 38.52 & 0.50 & 4   \\
56309.6904 & 38.82 & 0.10 & 4  & & & & 56703.5994 & 38.39 & 0.50 & 4   \\
56310.7357 & 38.85 & 0.10 & 4  & & & & 57410.7730 & 38.01 & 0.50 & 4   \\
56323.6397 & 39.19 & 0.10 & 4  & & & &  & & &  \\
56324.7170 & 38.85 & 0.10 & 4  & & & &  & & &  \\
56677.6733 & 38.85 & 0.10 & 4  & & & &  & & &  \\
57385.8825 & 38.90 & 0.10 & 4  & & & &  & & &  \\
\bottomrule\\
\end{tabular} 
\begin{tablenotes}
\scriptsize\vspace{-1em}
\item $^{\dagger}$ 1: MMT Digital Speedometer; 2: Tillinghast Reflector Digital Speedometer; 3: Wyeth Reflector Digital Speedometer; 4: Tillinghast Reflector Echelle Spectrograph
\end{tablenotes}
\end{table*}

\newpage

\section{BASTA model distributions}
\label{sec:basta_dist}

\begin{figure*}
    \centering
    \begin{subfigure}
        \centering
        \includegraphics[width=\textwidth]{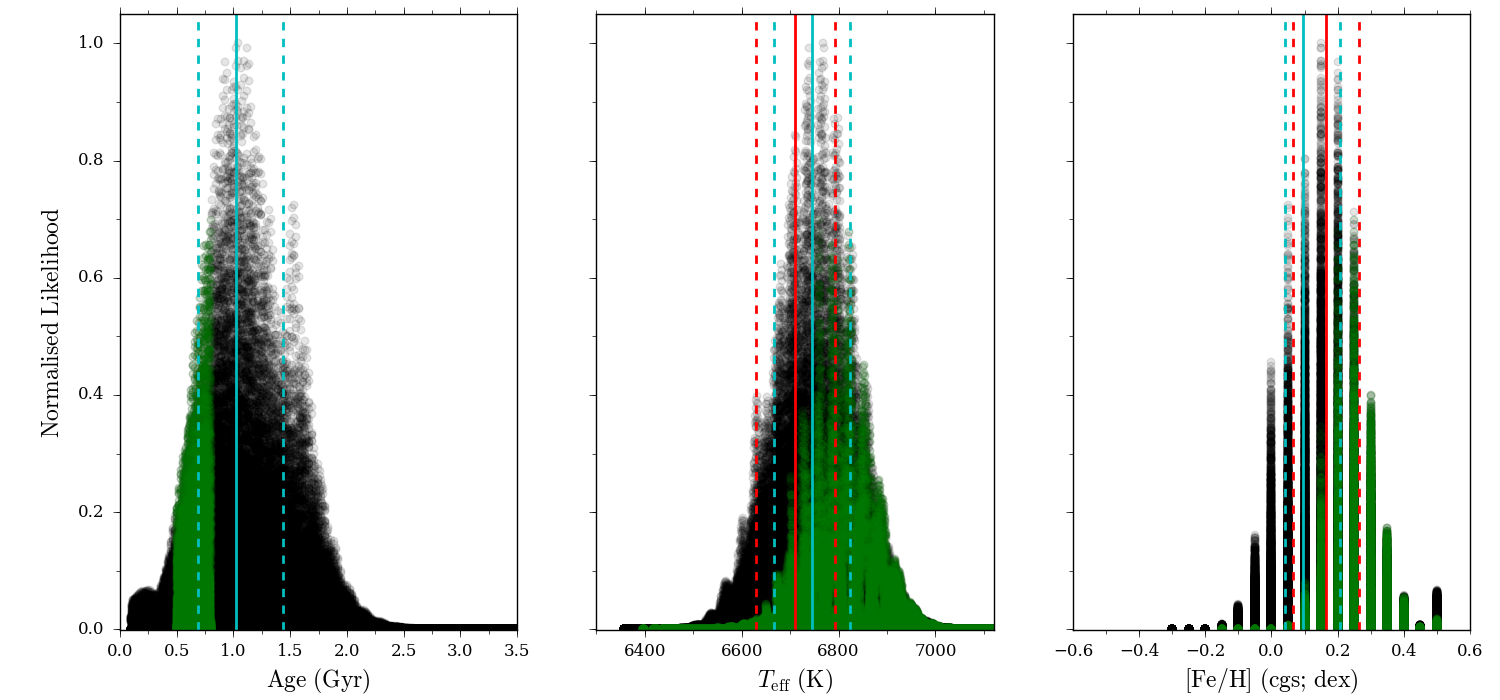}
    \end{subfigure}%
    ~ 
    \begin{subfigure}
        \centering
        \includegraphics[width=\textwidth]{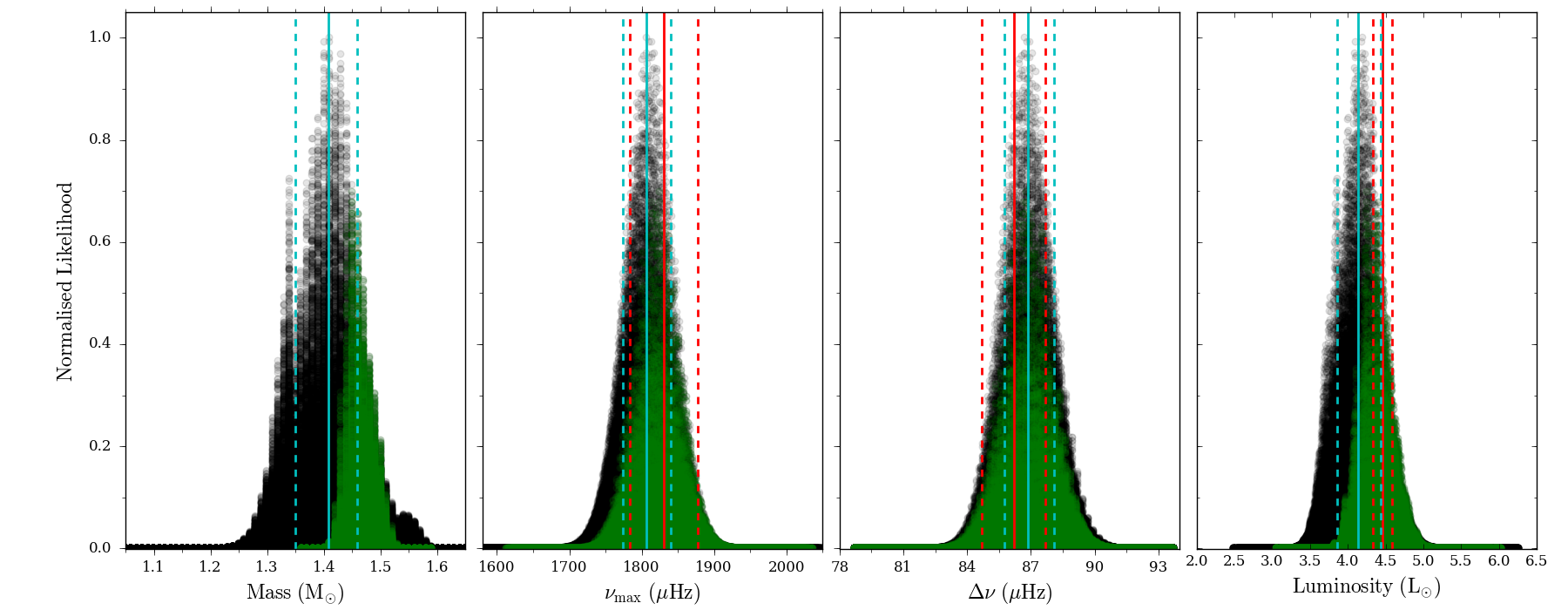}
    \end{subfigure}
    \caption{Posterior distributions from the GBM of \targ with BASTA. The different panels give the distributions for different model quantities of interest, with observed average seismic parameters and spectroscopic inputs from SPC indicated by vertical red lines, and final model values from BASTA indicated by vertical cyan lines; dashed lines indicate the $1-\sigma$ values on the parameters. For the luminosity the indicated value is calculated from the SPC \teff and the distance from the parallax of \citet[][]{2002A&A...381..446M}. The models shown in green are the ones falling in the age interval between $500-800$ Myr. In all cases likelihood values have been normalised to a maximum of 1.}
\label{fig:basta_dist}
\end{figure*}

\fref{fig:basta_dist} Presents the posterior distributions from the GBM of \targ with BASTA.
As seen from the models with an age in the interval between $500-800$ Myr (marked in green) a higher metallicity is preferred for a good reconciliation with isochrone-based ages. The higher \feh gives a corresponding increase in \teff and mass, and with it a decrease in age. It is also clear that the models that provide an age as expected for the Hyades still match the average seismic parameters for the star, indicating that these only contribute with a mild constraint on age.

\label{lastpage}
\end{document}